\documentclass[superscriptaddress, reprint, amssymb, amsmath, aip, jcp]{revtex4-1}
\usepackage{graphicx, color, url}

\newcommand{\beq}{\begin{equation}}
\newcommand{\eeq}{\end{equation}}

\renewcommand{\S}{S}
\newcommand{\Sfinal}{S_\mathrm{final}}
\newcommand{\s}{\sigma}

\renewcommand{\th}[1]{\theta^{(#1)}}
\newcommand{\Pcal}{\mathcal{P}}
\newcommand{\Fcal}{\mathcal{F}}
\newcommand{\Lcal}{\mathcal{L}}
\newcommand{\Tcal}{\mathcal{T}}
\newcommand{\Scal}{\mathcal{S}}
\newcommand{\Ucal}{\mathcal{U}}
\newcommand{\average}[1]{\left\langle #1 \right\rangle}
\newcommand{\Tmom}[1]{\average{\mathcal{T}^{(#1)}}}
\newcommand{\Lmom}[1]{\average{\mathcal{L}^{#1}}}
\newcommand{\Smom}[1]{\average{\mathcal{S}^{#1}}}
\newcommand{\Umom}[1]{\average{\mathcal{U}^{#1}}}

\newcommand{\me}[3]{\langle #1 | #2 | #3 \rangle}
\newcommand{\bra}[1]{\langle #1 |}
\newcommand{\ket}[1]{| #1 \rangle}
\newcommand{\innerprod}[2]{\langle #1 | #2 \rangle}
\newcommand{\Qb}{\mathbf{Q}}
\newcommand{\Wb}{\mathbf{W}}
\newcommand{\Qbtilde}{\tilde{\mathbf{Q}}} 
\newcommand{\Tb}{\mathbf{T}}
\newcommand{\Sb}{\mathbf{S}}
\newcommand{\Ub}{\mathbf{U}}
\newcommand{\Ob}{\mathbf{\Omega}}
\newcommand{\Thb}{\mathbf{\Theta}}
\newcommand{\Kb}{\mathbf{K}}
\newcommand{\Gb}{\mathbf{G}}
\newcommand{\Fb}{\mathbf{F}}

\newcommand{\zero}{\mathbf{0}}
\newcommand{\nmax}{n_\mathrm{max}}

\newcommand{\lbar}{\bar{\ell}}
\newcommand{\tbar}{\bar{t}}
\newcommand{\sbar}{\bar{s}}

\newcommand{\std}{\mathrm{std}}
\renewcommand{\c}{\mathrm{c}}
\newcommand{\cv}{\mathrm{cv}}

\newcommand{\bulk}{\mathrm{bulk}}
\newcommand{\edge}{\mathrm{edge}}
\newcommand{\backbone}{\mathrm{backbone}}
\newcommand{\tooth}{\mathrm{tooth}}
\newcommand{\macro}{\mathrm{macro}}
\newcommand{\micro}{\mathrm{micro}}

\newcommand{\red}[1]{#1}

\begin{document}


\title{Path statistics, memory, and coarse-graining of continuous-time random walks on networks}

\author{Michael Manhart}
\email{\texttt{mmanhart@fas.harvard.edu}}
\affiliation{Department of Physics and Astronomy, Rutgers University, Piscataway, NJ 08854, USA}
\affiliation{Department of Chemistry and Chemical Biology, Harvard University, Cambridge, MA 02138, USA}

\author{Willow Kion-Crosby}
\affiliation{Department of Physics and Astronomy, Rutgers University, Piscataway, NJ 08854, USA}

\author{Alexandre V. Morozov}
\email{\texttt{morozov@physics.rutgers.edu}}
\affiliation{Department of Physics and Astronomy, Rutgers University, Piscataway, NJ 08854, USA}

\date{\today}

\begin{abstract}
Continuous-time random walks (CTRWs) on discrete state spaces, ranging from regular lattices to complex networks, are ubiquitous across physics, chemistry, and biology.  Models with coarse-grained states, for example those employed in studies of molecular kinetics, and models with spatial disorder can give rise to memory and non-exponential distributions of waiting times and first-passage statistics.  However, existing methods for analyzing CTRWs on complex energy landscapes do not address these effects.  We therefore use statistical mechanics of the nonequilibrium path ensemble to characterize first-passage CTRWs on networks with arbitrary connectivity, energy landscape, and waiting time distributions.  Our approach is valuable for calculating higher moments (beyond the mean) of path length, time, and action, as well as statistics of any conservative or non-conservative force along a path.  For homogeneous networks we derive exact relations between length and time moments, quantifying the validity of approximating a continuous-time process with its discrete-time projection.  For more general models we obtain recursion relations, reminiscent of transfer matrix and exact enumeration techniques, to efficiently calculate path statistics numerically.  We have implemented our algorithm in PathMAN, a Python script that users can easily apply to their model of choice.  We demonstrate the algorithm on a few representative examples which underscore the importance of non-exponential distributions, memory, and coarse-graining in CTRWs.
\end{abstract}

\maketitle



\section{Introduction}

     We can model many dynamical systems in physics, chemistry, and biology as random walks on discrete state spaces or network structures.  For example, random walks can represent proteins folding on a coarse-grained network of conformational states~\cite{Noe2009, Lane2011}, particles diffusing in disordered, fractal-like media~\cite{benAvraham2000, Redner2001}, populations evolving in DNA or protein sequence space~\cite{Weinreich2006, Manhart2015}, and cells differentiating across epigenetic landscapes of regulatory states~\cite{Enver2009, Lang2014}.  One can also use random walks to probe the structure of empirical complex networks, such as protein-protein interaction networks or the World Wide Web~\cite{Albert2002, Gallos2007, Condamin2007}.  The central problem in these models is characterizing the statistical properties of paths taken by the system as it evolves from one state to another, especially for systems out of equilibrium.  This entails understanding not only the distribution of lengths and times for these paths, but also their distribution in the state space, which may reveal bottlenecks and indicate the diversity of intermediate pathways. 

     There is extensive literature for random walks on lattices~\cite{Montroll1965, Weiss1994, Redner2001}, fractals~\cite{Redner2001, benAvraham2000}, and random and scale-free networks~\cite{Albert2002, Bollt2005, Condamin2007} in the absence of an energy landscape or other objective function.  Much of this work has focused especially on the scaling behavior of first-passage times and the mean square displacement, the latter being important to identifying anomalous diffusion~\cite{benAvraham2000}.  More complex models, especially those with energy landscapes derived from empirical models or experimental data, generally require numerical approaches.  One such approach is transition path theory~\cite{Metzner2009}, which relies on numerical solutions of the backward equation.  This technique has been used to study Markov state models of molecular kinetics such as protein folding~\cite{Noe2009, Lane2011}.
     
     Standard transition path theory, however, is not applicable to general continuous-time random walks~\cite{Weiss1994} (CTRWs) where states may have non-exponential waiting time distributions, nor does it address the complete distribution of first-passage times beyond the mean.  These problems are important in many systems.  For example, molecular Markov state models require grouping large numbers of microscopic conformations of molecules into a small number of effective states~\cite{Prinz2011}; the stochastic dynamics are then analyzed on this effective model~\cite{Noe2009, Lane2011}.  However, this coarse-graining is known to lead to qualitative differences with the underlying microscopic dynamics~\cite{Prinz2011}.  In particular, the loss of information due to coarse-graining can lead to the appearance of memory, manifested as non-exponential waiting time distributions, in the coarse-grained states.  Indeed, there is evidence of non-exponential distributions of time in protein conformation dynamics~\cite{Sabelko1999, Yang2003} and enzyme kinetics~\cite{Flomenbom2005b, Reuveni2014}.  Non-exponential distributions can also arise from spatial disorder, as in glassy systems~\cite{Campbell1988, Angelani1998}.  Other linear algebra-based methods have been developed to treat general CTRWs~\cite{Hunter1969, Yao1985, Harrison2002}, but such methods are complicated and provide relatively little physical insight.
     
     \red{An alternative, more intuitive approach to CTRWs uses the path representation: statistical properties of CTRWs are decomposed into averages over the ensemble of all possible stochastic paths through state space. Some analytical results with this approach for arbitrary energy landscapes and waiting time distributions have been obtained, but only for 1D lattices, due to the difficulty of enumerating paths~\cite{Flomenbom2005a, Flomenbom2007a, Flomenbom2007b}.  On the other hand, path sampling methods~\cite{Harland2007} are able to treat arbitrary network topologies, but these methods have not been developed for non-exponential waiting time distributions, and in any case sampling is likely to be inefficient for calculating higher moments of path statistics, which are crucial when non-exponential distributions are expected.}
     
     Here we develop a generalized formalism for the path ensemble of a CTRW on a network of discrete states, regardless of their connectivity, energy landscape, or intermediate state waiting time distributions.
In Sec.~\ref{sec:theory} we use statistical mechanics of the nonequilibrium path ensemble for a CTRW to obtain expressions for arbitrary moments of path statistics including path length, time, action, and any conservative or nonconservative force along a path.  We use this formalism to deduce general relationships among the distributions of path time, length, and action, as well as several exact relationships for the case of homogeneous networks.
In Sec.~\ref{sec:algorithm} we derive recursion relations, reminiscent of transfer matrix and exact enumeration techniques, to efficiently calculate various path statistics numerically, including distributions of paths in the state space.  We have implemented our approach in a user-friendly Python script called PathMAN (Path Matrix Algorithm for Networks), freely available at \url{https://github.com/michaelmanhart/pathman}, that users can apply to their own models.
     
     In Sec.~\ref{sec:examples} we demonstrate the numerical algorithm on a few examples.  After illustrating some basic concepts on a simple 1D random walk, we apply our method to a 1D comb to show how coarse-graining can lead to the appearance of memory, in the form of non-exponential waiting time distributions.  We quantify the effect of the memory on the distribution of total path times.  We further demonstrate the effect of coarse-graining in a 2D double-well potential, from which we deduce some general properties of memory arising from coarse-graining.  Lastly, we use our method to show how spatial disorder can also lead to non-exponential distributions of path statistics in the 2D random barrier model.

\section{Distributions in the path ensemble}
\label{sec:theory}


\subsection{Continuous-time random walks and memory}

     Consider a stochastic process on a finite set $\S$ of $N$ states: the process makes discrete jumps between states with continuous-time waiting at each state in between jumps.  Such a process is known as a continuous-time random walk~\cite{Weiss1994} (CTRW), and it can describe many physical or biological systems, such as a protein traversing a coarse-grained network of conformations toward its folded state~\cite{Noe2009, Lane2011} or a particle traveling through a disordered material~\cite{benAvraham2000, Redner2001}.  The time the system waits in a state $\s$ before making a jump to $\s'$ is distributed according to $\psi(t|\s\to\s')$.  In many models this distribution depends only on the current state $\s$ and not on the destination $\s'$, so that $\psi(t|\s\to\s') = \psi(t|\s)$; such waiting time distributions are known as ``separable''~\cite{Haus1987}.  We will mostly assume separable distributions throughout this paper.  However, since non-separable distributions arise crucially in coarse-grained models, we will also briefly discuss how to extend our results to the non-separable case.  Let the raw moments of the waiting time distributions be denoted as

\beq
\th{n}(\s) = \int_0^\infty dt~ \psi(t|\s)~ t^n.
\label{eq:gen_time_moments}
\eeq
     
\noindent We assume every state $\s$ has at least one finite moment for $n>0$; the zeroth moment is always $\th{0}(\s) = 1$ by normalization. In the special case of a discrete time process, $\psi (t|\s) = \delta(t - \th{1}(\s))$ and the moments are $\th{n}(\s) = (\th{1}(\s))^n$.
     
     Given the system has finished waiting in $\s$ and makes a jump out, the probability of jumping to $\s'$ is given by the matrix element $\me{\s'}{\Qb}{\s}$, where $\Qb$ is an $N \times N$ matrix and $\ket{\s}$ denotes an $N$-dimensional vector with $1$ at the position corresponding to the state $\s$ and 0 everywhere else.  The jump probabilities out of each state $\s$ are therefore normalized according to $\sum_{\s'} \me{\s'}{\Qb}{\s} = 1$, \red{with $\me{\s}{\Qb}{\s} = 0$ by definition (since a jump must leave the current state).}  The matrix $\Qb$ imposes a network structure over the states in $\S$, with edges directed and weighted by the entries in $\Qb$.  We can think of the jump process alone as a discrete-time projection of the model, since it describes the system's dynamics if we integrate out the continuous waiting times.

     An ordinary Markov process is a special case of the above CTRW construction.  A continuous-time Markov process is typically defined by a rate matrix $\Wb$ such that in a small time interval $\Delta t$, the probability of making a jump $\s \to \s'$ is $\me{\s'}{\Wb}{\s} \Delta t$.  Therefore the probability of making the jump $\s \to \s'$, given that the system makes any jump out of $\s$ during $\Delta t$, is
    
\beq
\frac{\me{\s'}{\Wb}{\s} \Delta t}{\sum_{\s''} \me{\s''}{\Wb}{\s} \Delta t} = \frac{\me{\s'}{\Wb}{\s}}{\sum_{\s''} \me{\s''}{\Wb}{\s}} = \me{\s'}{\Qb}{\s},
\label{eq:Markov_jump_prob}
\eeq

\noindent which defines the relation between the Markov rate matrix $\Wb$ and the jump matrix $\Qb$.  The probability per unit $\Delta t$ of waiting time $t = M\Delta t$ in $\s$ and then making a jump out is given by

\beq
\frac{1}{\Delta t} \left(\sum_{\s'} \me{\s'}{\Wb}{\s} \Delta t \right) \left(1 - \sum_{\s'} \me{\s'}{\Wb}{\s} \Delta t \right)^M.
\label{eq:nojump_jump}
\eeq

\noindent The waiting time distribution $\psi(t|\s)$ is then the continuous limit of Eq.~\ref{eq:nojump_jump}:

\beq
\begin{split}
\psi(t|\s) & = \lim_{\Delta t \to 0} \left(\sum_{\s'} \me{\s'}{\Wb}{\s} \right) \left(1 - \sum_{\s'} \me{\s'}{\Wb}{\s} \Delta t \right)^{t/\Delta t} \\
& = \frac{1}{\th{1}(\s)} e^{-t/\th{1}(\s)}, \\
\end{split}
\label{eq:Markov_psi}
\eeq

\noindent where

\beq
\th{1}(\s) = \left( \sum_{\s'} \me{\s'}{\Wb}{\s} \right)^{-1}
\label{eq:Markov_wtm}
\eeq

\noindent is the mean waiting time in $\s$.  Hence waiting times in a Markov process always have an exponential distribution.  The higher moments of exponential waiting times are \red{completely determined by the mean}: $\th{n}(\s) = n! (\th{1}(\s))^n$.
    

     In general, processes with exponential distributions of times $p(t)$ are important because they are memoryless in the following sense: the probability of taking at least time $t$, given the process has already taken at least time $t_0$, is the same as taking at least $t$ in the first place.  That is, the system ``forgets'' the time it has already taken.  Mathematically this means that

\beq
\frac{P(t+t_0)}{P(t_0)} = P(t),
\label{eq:memoryless}
\eeq

\noindent where $P(t) = \int_t^\infty dt'~ p(t')$ is the complementary cumulative distribution function.  The only function satisfying Eq.~\ref{eq:memoryless} is a simple exponential $P(t) = e^{-t/\tau}$, from which it follows that $p(t) = \tau^{-1} e^{-t/\tau}$.  For waiting time distributions $\psi(t|\s)$, non-exponential functions are therefore indicative of memory within a state: how much longer the system tends to wait in that state depends on how long it has already waited.  In contrast to ordinary Markov models where $\psi(t|\s)$ is always exponential, models where $\psi(t|\s)$ may be non-exponential are sometimes known as semi-Markov processes. 



\subsection{The ensemble of first-passage paths}

     We approach CTRWs using the ensemble of first-passage paths~\cite{Flomenbom2005a, Sun2006, Flomenbom2007a, Flomenbom2007b, Harland2007, Manhart2013, Manhart2014}, which first reach a particular final state or a set of final states from some initial conditions.  We are interested in statistical properties of this ensemble such as its distributions in length, time, and space. 
In addition to situations where first-passage properties themselves are of interest, first-passage paths constitute fundamental building blocks of a stochastic process since the full propagator and steady state can in principle be derived from them~\cite{Redner2001}.

     Let $\Sfinal$ be the set of final states, which we will treat as absorbing ($\me{\s'}{\Qb}{\s} = 0$ for all $\s \in \Sfinal$ and $\s' \in \S$) so that the first-passage condition is satisfied. 
Define a path $\varphi$ of length $\ell$ to be an ordered sequence of $\ell+1$ states: $\varphi = \{\s_0, \s_1, \ldots, \s_\ell\}$.  Denote the probability distribution over initial states as $\pi_0(\s)$.  Then the probability density of starting in a state $\s_0$ and completing the path $\varphi$ at exactly time $t$ is given by

\begin{multline}
\Pcal[\varphi, t] = \pi_0(\s_0) \left(\prod_{i=0}^{\ell-1} \me{\s_{i+1}}{\Qb}{\s_i} \right) \\ 
\times \left[ \int_0^\infty dt_0~ \psi(t_0|\s_0)~ \int_0^\infty dt_1~ \psi(t_1|\s_1)~ \cdots \right. \\
\left. \int_0^\infty dt_{\ell-1}~ \psi(t_{\ell-1}|\s_{\ell-1})~ \delta\left( t - \sum_{i = 0}^{\ell-1} t_i \right) \right],
\label{eq:time_dependent_path_prob}
\end{multline}

\noindent where $t_0, t_1, \ldots, t_{\ell-1}$ are the intermediate waiting times and $\delta$ is the Dirac delta function. The probability of completing the path $\varphi$ irrespective of how much time it takes is then

\beq
\Pcal[\varphi] = \int_0^\infty dt~ \Pcal[\varphi,t] = \pi_0(\s_0) \prod_{i=0}^{\ell-1} \me{\s_{i+1}}{\Qb}{\s_i}.
\label{eq:path_prob}
\eeq

     The time-independent path probability $\Pcal[\varphi]$ is convenient because we can express many path statistics of interest as averages over this distribution, analogous to averages over the Boltzmann distribution in ordinary statistical mechanics~\cite{Harland2007, Manhart2013, Manhart2014}.  For example, let $\Fcal[\varphi]$ be a functional that measures some property of the path $\varphi$.  We use angle brackets to denote the average of this quantity over the path ensemble:
     
\beq
\average{\Fcal} = \sum_\varphi \Pcal[\varphi] \Fcal[\varphi],
\label{eq:path_average_def}
\eeq

\noindent where the sum is over all first-passage paths $\varphi$ of any length ending at states in $\Sfinal$.  Note that the partition function of the first-passage path ensemble, $\sum_\varphi \Pcal[\varphi]$, always equals 1, since the process must reach one of the final states eventually.  In this manner we can calculate \emph{nonequilibrium} (first-passage) properties of the system as \emph{equilibrium} properties of the path ensemble, which is time-independent by construction.


\subsection{Distribution of path lengths}

     The simplest path property is its length $\Lcal[\varphi]$, i.e., the discrete number of jumps along the path.  The mean path length is then 

\beq
\average{\Lcal} = \sum_\varphi \Pcal[\varphi] \Lcal[\varphi].
\eeq

\noindent Functionals for the higher moments of path length are simply powers of the length functional,

\beq
\Lmom{n} = \sum_\varphi \Pcal[\varphi] \left(\Lcal[\varphi]\right)^n,
\label{eq:length_moments}
\eeq

\noindent and the path length probability distribution is 

\beq
\rho(\ell) = \average{\delta_{\ell,\Lcal}},
\label{eq:rho_def}
\eeq

\noindent where $\delta$ is the Kronecker delta.  Note that the distribution of path lengths depends only on the jump matrix $\Qb$ and not on the waiting time distributions $\psi(t|\s)$, \red{and hence it can be thought to characterize the discrete-time projection of the underlying continuous-time stochastic process.}  In Appendix~\ref{sec:asymptotic_rho} we show that the distribution of path lengths $\rho(\ell)$ is typically exponential asymptotically:

\beq
\rho(\ell) \sim e^{-\alpha\ell/\lbar},
\label{eq:asymptotic_rho}
\eeq

\noindent where $\alpha$ is a constant of order 1 and $\lbar = \Lmom{}$ is the mean path length.

%
%


\subsection{Distribution of path times}

     In contrast to the discrete length of a path, there is also the continuous time of the path that accounts for the variable waiting times at the intermediate states.  The distribution of total path times (first-passage time distribution) is


\beq
f(t) = \sum_\varphi \Pcal[\varphi, t].
\label{eq:path_time_distribution}
\eeq

\noindent Unlike the path length distribution, the path time distribution depends on \emph{both} the jump matrix $\Qb$ as well as the waiting time distributions $\psi(t|\s)$.  We cannot evaluate $f(t)$ for arbitrary waiting time distributions $\psi(t|\s)$; however, we can express its moments as simple averages over the time-independent path ensemble using path functionals (cf. Eq.~\ref{eq:path_average_def}).  That is, using Eqs.~\ref{eq:time_dependent_path_prob} and~\ref{eq:path_time_distribution}, we obtain

\beq
\begin{split}
\int_0^\infty dt~ f(t)~ t^n & = \int_0^\infty dt~ t^n \sum_\varphi \Pcal[\varphi, t] \\
& = \sum_\varphi \Pcal[\varphi] \Tcal^{(n)}[\varphi] \\
& = \Tmom{n}, \\
\end{split}
\label{eq:time_moment}
\eeq

\noindent where the functional for the $n$th moment of path time is

\beq
\begin{split}
\Tcal^{(n)}[\varphi] & = \int_0^\infty dt_0~ \psi(t_0|\s_0)~ \int_0^\infty dt_1~ \psi(t_1|\s_1)~ \cdots \\
& \qquad \int_0^\infty dt_{\ell-1}~ \psi(t_{\ell-1}|\s_{\ell-1})~
\left( \sum_{i = 0}^{\ell-1} t_i \right)^n \\
& = \sum_{j_0, j_1, \ldots, j_{\ell-1}} {n \choose j_0, j_1, \ldots, j_{\ell-1}} \\
& \qquad \times \th{j_0}(\s_0) \th{j_1}(\s_1) \cdots \th{j_{\ell-1}}(\s_{\ell-1}).
\end{split}
\label{eq:time_functional}
\eeq

\noindent Each summation in the multinomial expansion is from $0$ to $n$ subject to the constraint $j_0 + j_1 + \cdots + j_{\ell-1} = n$.  For example, the first few moments are

\beq
\begin{split}
\mathcal{T}^{(1)}[\varphi] = & \sum_{i=0}^{\ell-1} \th{1}(\s_i), \\
\mathcal{T}^{(2)}[\varphi] = & \sum_{i=0}^{\ell-1} \th{2}(\s_i) + 2\sum_{i<j} \th{1}(\s_i) \th{1}(\s_j),\\
\mathcal{T}^{(3)}[\varphi] = & \sum_{i=0}^{\ell-1} \th{3}(\s_i) \\ 
& + 3\sum_{i<j} \left(\th{1}(\s_i) \th{2}(\s_j) + \th{2}(\s_i) \th{1}(\s_j)\right) \\
& + 6\sum_{i<j<k} \th{1}(\s_i) \th{1}(\s_j) \th{1}(\s_k). \\
\end{split}
\label{eq:low_order_functionals}
\eeq

\noindent Note that Eq.~\ref{eq:time_functional} implies that if any accessible intermediate state has a divergent waiting time moment of order $n$, then all path time moments of order $n$ and higher must be divergent as well.


\subsection{Path action and a general class of path functionals}

     For many systems it is important to determine whether their dynamics are highly predictable or highly stochastic; that is, whether the system is likely to take one of a few high-probability paths every time, or whether there is a more diverse ensemble of paths with similar probabilities.
One way to quantify this notion uses the path action, defined as

\beq
\Scal[\varphi] = - \sum_{i=0}^{\ell-1} \log \me{\s_{i+1}}{\Qb}{\s_i},
\eeq

\noindent so that path probability is $\Pcal[\varphi] = \pi_0(\s_0) e^{-\Scal[\varphi]}$.
As in classical and quantum mechanics, paths with minimum action dominate while paths of large action are suppressed.  Note that like path lengths, action depends only on the jump probabilities and not on the waiting time distributions.

     The mean path action is the Shannon entropy of the path distribution~\cite{Filyukov1967} (we ignore the path-independent $\log \pi_0(\s_0)$ contribution from the initial condition):

\beq
\Smom{} = - \sum_\varphi \Pcal[\varphi] \log \Pcal[\varphi].
\eeq

\noindent This is consistent with the idea that the path action distribution tells us about the diversity of paths in the ensemble: low entropy (small mean action) means that a few paths with large probability dominate the process, while large entropy (large mean action) means that a diverse collection of low-probability paths contribute.  The distribution of actions around this mean may be non-trivial, however.  For instance, even if the mean action is large, the variance around it could either be small (the system must traverse one of the low-probability paths) or large (the system may traverse paths with a wide range of probabilities).  We can characterize the action distribution by considering its higher moments.  The functional for the $n$th moment of path action is
     
\beq
\begin{split}
\left(\Scal[\varphi]\right)^n & = \left(- \sum_{i=0}^{\ell-1} \log \me{\s_{i+1}}{\Qb}{\s_i} \right)^n \\
& = \sum_{j_0, j_1, \ldots, j_{\ell-1}} {n \choose j_0, j_1, \ldots, j_{\ell-1}} \\
 & \qquad \times \prod_{i=0}^{\ell-1} (-\log\me{\s_{i+1}}{\Qb}{\s_i})^{j_i}, 
\end{split}
\label{eq:action_functional}
\eeq

\noindent so the total moments of the path action distribution are

\beq
\Smom{n} = \sum_\varphi \Pcal[\varphi] \left(\Scal[\varphi]\right)^n.
\eeq

\noindent 
     
     The action functionals (Eq.~\ref{eq:action_functional}) share a similar multinomial form with the time functionals (Eq.~\ref{eq:time_functional}).  This leads us to consider a more general class of path functionals with this form.  Consider a path functional $\Ucal$ that sums some property over jumps in a path (or edges in the network), so that for a path $\varphi$ of length $\ell$, 
     
\beq
\Ucal[\varphi] = \sum_{i=0}^{\ell-1} U(\s_{i+1}, \s_i).
\label{eq:general_functional}
\eeq

\noindent In the case of action, $U(\s_{i+1}, \s_i) = -\log \me{\s_{i+1}}{\Qb}{\s_i}$.  Equation~\ref{eq:general_functional} is a discretized line integral along the path $\varphi$, which suggests thinking of $U$ as representing a force acting on the random walker as it traverses a path.  The statistics of such forces over paths are especially interesting when the force is nonconservative, i.e., the line integral $\Ucal[\varphi]$ depends on the whole path $\varphi$ and not just on the end points.  Non-transitive landscapes or non-gradient forces with this property have been considered in evolutionary theory~\cite{Mustonen2010} and biochemical networks~\cite{Wang2008}.
However, even for conservative forces, the distribution of the line integral $\Ucal[\varphi]$ may be non-trivial over the path ensemble if there are multiple initial and final states.  The moment functionals for any such quantity are again of the multinomial form:

\begin{multline}
(\Ucal[\varphi])^n = \\ 
\sum_{j_0,j_1,\ldots,j_{\ell-1}} {n \choose j_0, j_1, \ldots, j_{\ell-1}} \prod_{i=0}^{\ell-1} \left(U(\s_{i+1}, \s_i)\right)^{j_i}. 
\end{multline}

\noindent This suggests that methods for calculating time or action moments can be applied to any path property in this general class.


\subsection{Path statistics on a homogeneous network}
\label{sec:homogeneous}

     We now consider these statistics of path length, time, and action in the simple case of a network with homogeneous properties.  We first assume that the waiting time distributions $\psi(t|\s) = \psi(t)$ are identical for all states, with raw moments $\th{n}$ and cumulant moments $\th{n}_\c$.  We do not assume anything about the jump matrix $\Qb$ (i.e., the network connectivity).  In Appendix~\ref{sec:generating_functions} we derive an exact relation between path time and length moments for arbitrary $\psi(t)$:

\beq
\Tmom{n} = \sum_{k=1}^n \Lmom{k} B_{n,k}\left(\th{1}_\c, \th{2}_\c, \ldots, \th{n-k+1}_\c\right),
\label{eq:times_vs_lengths}
\eeq

\noindent where $B_{n,k}$ are the partial Bell polynomials~\cite{Comtet1974}. For example, the first few moments are 

\beq
\begin{split}
\Tmom{1} & = \th{1}_\c \Lmom{}, \\
\Tmom{2} & = \left(\th{1}_\c\right)^2 \Lmom{2} + \th{2}_\c \Lmom{}, \\
\Tmom{3} & = \left(\th{1}_\c\right)^3 \Lmom{3} + 3 \th{1}_\c \th{2}_\c \Lmom{2} + \th{3}_\c \Lmom{}. \\
\end{split}
\label{eq:times_vs_lengths_example}
\eeq

\noindent Note that the $n$th time moment depends on all length moments up to $n$.  Equation~\ref{eq:times_vs_lengths} holds for the cumulants $\Tmom{n}_\c$ and $\Lmom{n}_\c$ as well (Appendix~\ref{sec:generating_functions}).  \red{In Appendix~\ref{sec:central_limit_theorem} we present an alternative argument for the first two moments of Eq.~\ref{eq:times_vs_lengths_example} using the central limit theorem, and in Appendix~\ref{sec:time_length_moments_exponential} we study the special case when $\psi(t)$ is exponential.}

     As these results show, we can think of path time as a convolution between path length and the intermediate waiting times: the variation in total path times arises from both variation in path lengths as well as variation in the waiting times.  If $\psi(t)$ is a delta function (discrete-time process), then $\th{n}_\c = 0$ for $n>1$ (no variation in waiting times), and $\Tmom{n} = (\th{1})^n \Lmom{n}$ exactly: path lengths and times are identical up to an overall scale.  This is consistent with our previous notion that the path length distribution fully describes the discrete-time projection of the process.  However, even with continuous-time distributions $\psi(t)$, the approximation $\Tmom{n} \approx (\th{1})^n \Lmom{n}$, and therefore the approximate equivalence of the discrete- and continuous-time processes, may still hold if the waiting times are not too broadly dispersed (so the higher moments of $\psi(t)$ are not too large).  We can make this observation more quantitative by expanding Eq.~\ref{eq:times_vs_lengths} as
     
\beq
\begin{split}
\Tmom{n} & = \left(\th{1}_\c\right)^n \Lmom{n} \\
& \qquad \times \left[ 1 + \frac{\Lmom{n-1}}{\Lmom{n}} \frac{B_{n,n-1}\left(\th{1}_\c, \th{2}_\c\right)}{\left(\th{1}_\c\right)^n} + \cdots \right] \\
& = \left(\th{1}_\c\right)^n \Lmom{n} \\
& \qquad \times \left[ 1 + {n \choose 2} \left(\th{\cv}\right)^2 \frac{\Lmom{n-1}}{\Lmom{n}} + \cdots \right], \\
\end{split}
\label{eq:time_moment_approx}
\eeq
     
\noindent where 

\beq
\th{\cv} = \frac{\sqrt{\th{2}_\c}}{\th{1}}
\eeq

\noindent is the waiting time coefficient of variation (CV), i.e., the standard deviation divided by the mean.  The CV measures the relative dispersion of a distribution; it always equals 1 for exponential distributions.  As with Eq.~\ref{eq:times_vs_lengths}, Eq.~\ref{eq:time_moment_approx} holds for the cumulants $\Lmom{n}_\c$ and $\Tmom{n}_\c$ as well.

     Equation~\ref{eq:time_moment_approx} implies that path length and time moments will be approximately proportional, and hence the whole distributions should be similar, if

\beq
\left(\th{\cv}\right)^2 \frac{\Lmom{n-1}}{\Lmom{n}} \ll 1.
\label{eq:correction1}
\eeq

\noindent The quantity $\Lmom{n-1}/\Lmom{n}$ is typically of the order of the inverse mean path length $\Lmom{} = \lbar$; \red{in particular this is true when path lengths have an exponential distribution, which is generally the case asymptotically (Appendix~\ref{sec:asymptotic_rho}).}  An important exception, however, is if lengths have a Poisson distribution, so that $\Lmom{n-1}_\c/\Lmom{n}_\c = 1$ in the cumulant version of Eq.~\ref{eq:time_moment_approx}.  Apart from this special case, the condition of Eq.~\ref{eq:correction1} is equivalent to

\beq
\left(\th{\cv}\right)^2 \ll \lbar,
\label{eq:correction2}
\eeq

\noindent that is, the waiting time distribution must be sufficiently narrow compared to the mean path length.  In many cases we expect this to hold, since $\th{\cv} \sim 1$ for exponential-like waiting time distributions and the mean path length $\lbar$ is usually very large.  We will investigate the validity of this condition in later examples.






     We also consider path action on a homogeneous network.  Path action depends only on the jump matrix $\Qb$ and not on the waiting time distributions $\psi(t|\s)$, so as a simple example we take all states in $\S$ to have the same number $\gamma$ of outgoing jumps (nearest neighbors on the network) and all such jumps to have equal probability $\gamma^{-1}$.  Therefore the probability of a path is $\Pcal[\varphi] = \gamma^{-\Lcal[\varphi]}$, and the action is $\Scal[\varphi] = \Lcal[\varphi] \log \gamma$.  This means that the distribution of path actions is exactly equivalent to that of path lengths (rescaled by a factor of $\log\gamma$), and the moments are

\beq
\Smom{n} = \Lmom{n} \log^n \gamma.
\label{eq:homogeneous_action}
\eeq

\noindent Since path lengths typically have an exponential distribution asymptotically (Eq.~\ref{eq:asymptotic_rho}, Appendix~\ref{sec:asymptotic_rho}), path action will therefore also be asymptotically exponential as well, with mean $\lbar\log\gamma$. 



\section{Matrix formulation and numerical algorithm}
\label{sec:algorithm}

     Besides gaining general insights into the relationships between distributions of path lengths, times, and actions, the path ensemble formalism is convenient because we can efficiently calculate many ensemble averages using recursion relations that implicitly sum over all paths.  We now derive these relations and show how to implement them numerically.


\subsection{Recursion relations}

     We reformulate the problem in terms of matrices to express the sums over paths more explicitly.  Let $\Tb^{(n)}_{\ell}$ be an $N \times N$ matrix ($N$ is the number of states in $\S$) such that the matrix element $\me{\s'}{\Tb^{(n)}_{\ell}}{\s}$ is the $n$th time moment of all paths of exactly length $\ell$ from $\s$ to $\s'$.  In particular, the zeroth-order matrix $\Tb^{(0)}_{\ell}$ gives the total probability of all paths going from $\s$ to $\s'$ in exactly $\ell$ jumps. The initial condition is

\beq
\Tb^{(n)}_{0} = \delta_{n,0} \mathbf{1},
\label{eq:recursion_initial}
\eeq

\noindent where $\mathbf{1}$ is an $N \times N$ identity matrix.  If our path ensemble is the set of first-passage paths to final states $\Sfinal$ with an initial distribution vector $\ket{\pi_0} = \sum_{\s} \pi_0(\s) \ket{\s}$, then 

\beq
\sum_{\s\in\Sfinal} \me{\s}{\Tb^{(n)}_\ell}{\pi_0} = \sum_\varphi \Pcal[\varphi] \Tcal^{(n)}[\varphi] \delta_{\ell,\Lcal[\varphi]}
\eeq

\noindent is the $n$th time moment for all paths of exactly length $\ell$, and 

\beq
\sum_{\ell=0}^\infty \sum_{\s\in\Sfinal} \me{\s}{\Tb^{(n)}_\ell}{\pi_0} = \Tmom{n}
\label{eq:recursion_to_time_moments}
\eeq

\noindent is the moment averaged over paths of all lengths.  \red{This expression illustrates how to express the previous path ensemble averages in the matrix formulation.}

     The key advantage of the $\Tb^{(n)}_{\ell}$ matrices is that they obey the following recursion relation (Appendix~\ref{sec:recursion_derivation}):

\beq
\Tb^{(n)}_{\ell} = \Qb \sum_{j=0}^n {n \choose j} \Thb^{(j)} \Tb^{(n-j)}_{\ell-1},
\label{eq:time_recursion}
\eeq

\noindent where $\Thb^{(n)}$ is an $N \times N$ matrix with waiting time moments for each state along the diagonal:

\beq
\me{\s'}{\Thb^{(n)}}{\s} = \delta_{\s',\s} \th{n}(\s).
\label{eq:Thetaj_def}
\eeq

\noindent Appendix~\ref{sec:recursion_derivation} also shows how this recursion relation for the path time moments generalizes to the case of non-separable waiting time distributions $\psi(t|\s\to\s')$.  For the total probability ($n=0$), the recursion relation of Eq.~\ref{eq:time_recursion} is simply multiplication by the jump matrix: $\Tb^{(0)}_{\ell} = \Qb \Tb^{(0)}_{\ell-1}$, since the total probability of going from one state to another in exactly $\ell$ jumps must be given by the product of the jump matrices $\Tb^{(0)}_{\ell} = \Qb^\ell$.

     Owing to the similar multinomial form of their path functionals (compare Eqs.~\ref{eq:time_functional} and~\ref{eq:action_functional}), the path action moments obey a similar recursion relation.  Define $\me{\s'}{\Sb^{(n)}_\ell}{\s}$ to be the $n$th action moment of all paths of length $\ell$ from $\s$ to $\s'$, so that

\beq
\sum_{\ell=0}^\infty \sum_{\s\in\Sfinal} \me{\s}{\Sb^{(n)}_\ell}{\pi_0} = \Smom{n}.
\label{eq:recursion_to_action_moments}
\eeq

\noindent In Appendix~\ref{sec:recursion_derivation} we show that these matrices obey the recursion relation

\beq
\Sb^{(n)}_\ell = \sum_{j=0}^n {n \choose j} \Qbtilde^{(j)} \Sb^{(n-j)}_{\ell-1},
\label{eq:action_recursion}
\eeq

\noindent where the matrix $\Qbtilde^{(j)}$ is defined so that

\beq
\me{\s'}{\Qbtilde^{(j)}}{\s} = \me{\s'}{\Qb}{\s} (-\log \me{\s'}{\Qb}{\s})^j.
\label{eq:Qtilde_def}
\eeq

\noindent In fact, if $\Ub^{(n)}_\ell$ is the matrix such that

\beq
\sum_{\ell=0}^\infty \sum_{\s\in\Sfinal} \me{\s}{\Ub^{(n)}_\ell}{\pi_0} = \Umom{n}
\label{eq:general_functional_moments}
\eeq

\noindent for any path functional $\Ucal$ in Eq.~\ref{eq:general_functional}, it obeys the recursion relation

\beq
\Ub^{(n)}_\ell = \sum_{j=0}^n {n \choose j} \Ob^{(j)} \Ub^{(n-j)}_{\ell-1},
\label{eq:general_recursion}
\eeq

\noindent where $\me{\s'}{\Ob^{(j)}}{\s} = \me{\s'}{\Qb}{\s} (U(\s', \s))^j$ (Appendix~\ref{sec:recursion_derivation}).  Therefore recursion relations of this form extend to a wide class of path statistics.


\subsection{Transfer matrices}
\label{sec:transfer_matrices}

     To calculate the $n$th moment of time or action, we must carry out the recursion relation of Eq.~\ref{eq:time_recursion} or~\ref{eq:action_recursion} for all moments up to $n$.  We can unify all these steps into a single transfer matrix operation convenient for numerical use.  Let $\nmax$ be the maximum moment of interest.  Define the $N(\nmax+1)$-dimensional column vector $\ket{\tau(\ell)}$ as a concatenation of $\Tb^{(n)}_{\ell} \ket{\pi_0}$ for all $n \in \{0,1,\ldots,\nmax\}$:

\beq
\ket{\tau(\ell)}  = 
\left[
\begin{array}{c}
\Tb^{(0)}_{\ell} \ket{\pi_0} \\
\Tb^{(1)}_{\ell} \ket{\pi_0} \\
\vdots \\
\Tb^{(\nmax)}_{\ell} \ket{\pi_0} \\
\end{array}
\right].
\label{eq:tau_def}
\eeq

\noindent 
Define the basis vectors $\ket{\s,n}$ for $\s \in \S$ and $n \in \{0,1,\ldots,\nmax\}$ so that the $(\s,n)$ entry of $\ket{\tau(\ell)}$ is the $n$th time moment at state $\s$ at the $\ell$th jump: $\innerprod{\s,n}{\tau(\ell)} = \me{\s}{\Tb^{(n)}_\ell}{\pi_0}$.  We similarly define the action vector

\beq
\ket{\eta(\ell)} = 
\left[
\begin{array}{c}
\Sb^{(0)}_{\ell} \ket{\pi_0} \\
\Sb^{(1)}_{\ell} \ket{\pi_0} \\
\vdots \\
\Sb^{(\nmax)}_{\ell} \ket{\pi_0} \\
\end{array}
\right].
\label{eq:eta_def}
\eeq

     Now define the $N(\nmax+1) \times N(\nmax+1)$ matrices

\begin{widetext}
\beq
\begin{split}
\Kb = &
\left[
\begin{array}{ccccc}
{0 \choose 0} \Qb\Thb^{(0)} & \zero & \zero & \cdots & \zero \\
{1 \choose 1} \Qb\Thb^{(1)} & {1 \choose 0} \Qb\Thb^{(0)} & \zero & \cdots & \zero \\
{2 \choose 2} \Qb\Thb^{(2)} & {2 \choose 1} \Qb\Thb^{(1)} & {2 \choose 0} \Qb\Thb^{(0)} & \cdots & \zero \\
\vdots & \vdots & \vdots & \ddots & \vdots \\
{\nmax \choose \nmax} \Qb\Thb^{(\nmax)} & {\nmax\choose \nmax-1} \Qb\Thb^{(\nmax-1)} & {\nmax \choose \nmax-2} \Qb\Thb^{(\nmax-2)} & \cdots & {\nmax \choose 0} \Qb\Thb^{(0)} \\
\end{array}
\right], \\
& \\
\Gb = & 
\left[
\begin{array}{ccccc}
{0 \choose 0} \Qbtilde^{(0)} & \zero & \zero & \cdots & \zero \\
{1 \choose 1} \Qbtilde^{(1)} & {1 \choose 0} \Qbtilde^{(0)} & \zero & \cdots & \zero \\
{2 \choose 2} \Qbtilde^{(2)} & {2 \choose 1} \Qbtilde^{(1)} & {2 \choose 0} \Qbtilde^{(0)} & \cdots & \zero \\
\vdots & \vdots & \vdots & \ddots & \vdots \\
{\nmax \choose \nmax} \Qbtilde^{(\nmax)} & {\nmax \choose \nmax-1} \Qbtilde^{(\nmax-1)} & {\nmax \choose \nmax-2} \Qbtilde^{(\nmax-2)} & \cdots & {\nmax \choose 0} \Qbtilde^{(0)} \\
\end{array}
\right], \\
\label{eq:Kb_Gb_def}
\end{split}
\eeq
\end{widetext}

\noindent where each $\zero$ is an $N \times N$ zero matrix.  We can express the recursion relations of Eqs.~\ref{eq:time_recursion} and~\ref{eq:action_recursion} for all $n \in \{0,1,\ldots,\nmax\}$ as 

\beq
\ket{\tau(\ell)} = \Kb \ket{\tau(\ell-1)}, \quad \ket{\eta(\ell)} = \Gb \ket{\eta(\ell-1)}.
\eeq

\noindent These recursion relations have the solutions

\beq
\ket{\tau(\ell)} = \Kb^\ell \ket{\tau(0)}, \quad \ket{\eta(\ell)} = \Gb^\ell \ket{\eta(0)},
\eeq

\noindent where the initial conditions are

\beq
\ket{\tau(0)} = \ket{\eta(0)} = 
\left[
\begin{array}{c}
\ket{\pi_0} \\
\zero \\
\vdots \\
\zero \\
\end{array}
\right].
\label{eq:tau_eta_initial}
\eeq

\noindent Here each $\zero$ represents a zero column vector of length $N$. We can think of $\Kb$ and $\Gb$ as transfer matrices that iteratively generate sums over the path ensemble to calculate moments.  This is analogous to transfer matrices in spin systems that generate the sums over spin configurations to calculate the partition function~\cite{Yeomans1992}.  The zeroth-order version of this formalism, which simply calculates powers of the jump matrix $\Qb$, is equivalent to the exact enumeration method for discrete-time random walks~\cite{Majid1984, benAvraham2000}.

     We can obtain most path statistics of interest by various matrix and inner products on these vectors.  Define the cumulative moment vectors
     
\beq
\ket{\tau} = \sum_{\ell=0}^\infty \ket{\tau(\ell)}, \quad \ket{\eta} = \sum_{\ell=0}^\infty \ket{\eta(\ell)}.
\label{eq:cumulative_vectors}
\eeq

\noindent Elements of these vectors are (using Eqs.~\ref{eq:tau_def} and~\ref{eq:eta_def}) 

\beq
\begin{split}
\innerprod{\s,n}{\tau} & = \sum_{\ell=0}^\infty \me{\s}{\Tb^{(n)}_\ell}{\pi_0}, \\
\innerprod{\s,n}{\eta} & = \sum_{\ell=0}^\infty \me{\s}{\Sb^{(n)}_\ell}{\pi_0}.
\end{split}
\eeq

\noindent These represent the total $n$th moments of time and action for all paths through each state $\s$, but weighed by the number of visits to that state since the sum over $\ell$ counts a path's contribution each time it visits $\s$.  In the case of $n = 0$,

\beq
\innerprod{\s,0}{\tau} = \innerprod{\s,0}{\eta} = \sum_{\ell=0}^\infty \me{\s}{\Qb^\ell}{\pi_0},
\eeq

\noindent since $\Tb^{(0)}_\ell = \Sb^{(0)}_\ell = \Qb^\ell$ (Eqs.~\ref{eq:time_recursion} and~\ref{eq:action_recursion}).  This is actually the average number of visits $v(\s)$ to a state $\s$, since the probability of a path is counted each time it visits $\s$.  For an intermediate state $\s$, multiplying the mean number of visits $v(\s)$ by the mean waiting time $\th{1}(\s)$ gives the average time spent in $\s$.  When $\s$ is a final state in $\Sfinal$, the random walk can only visit it once (if it absorbs at that final state) or zero times (if it absorbs at a different final state), and thus the average number of visits $v(\s)$ equals the probability of reaching that final state $\s$ (commitment probability).  

     If there are multiple final states in $\Sfinal$, we often wish to sum path statistics over all of them.  Define the $N$-dimensional row vector $\bra{\mathrm{final}} = \sum_{\s \in \Sfinal} \bra{\s}$ (with 1 at the position for each final state and 0 everywhere else) and the $(\nmax+1) \times N(\nmax+1)$ matrix
     
\beq
\Fb = 
\left[
\begin{array}{cccc}
\bra{\mathrm{final}} & \zero & \cdots & \zero \\
\zero & \bra{\mathrm{final}} & \cdots & \zero \\
\vdots & \vdots & \ddots & \vdots \\
\zero & \zero & \cdots & \bra{\mathrm{final}} \\
\end{array}
\right],
\label{eq:Fb_def}
\eeq

\noindent where each $\zero$ is a zero row vector of length $N$.  Multiplying this matrix on a corresponding vector will sum over all final states for each moment, leaving an $(\nmax+1)$-dimensional vector with the total moments.  For example,

\beq
\begin{split}
\Fb \ket{\tau(\ell)} & = 
\left[
\begin{array}{c}
\me{\mathrm{final}}{\Tb^{(0)}_\ell}{\pi_0}  \\
\me{\mathrm{final}}{\Tb^{(1)}_\ell}{\pi_0} \\
\vdots \\
\me{\mathrm{final}}{\Tb^{(\nmax)}_\ell}{\pi_0} \\
\end{array}
\right] \\
& = 
\left[
\begin{array}{c}
\tbar^{(0)}(\ell) \\
\tbar^{(1)}(\ell) \\
\vdots \\
\tbar^{(\nmax)}(\ell) \\
\end{array}
\right] \\
& = \ket{\tbar(\ell)},
\end{split}
\eeq

\noindent where we use the shorthand $\tbar^{(n)}(\ell)$ for the total $n$th time moment absorbed at the $\ell$th jump. Note that $\tbar^{(0)}(\ell) = \rho(\ell)$ is the probability of reaching any of the final states in exactly $\ell$ jumps.  Thus this method allows us to calculate the entire path length distribution.  On the cumulative time vector $\ket{\tau}$, $\Fb$ returns the total time moments:

\beq
\Fb \ket{\tau} = 
\left[
\begin{array}{c}
\sum_{\ell=0}^\infty \me{\mathrm{final}}{\Tb^{(0)}_\ell}{\pi_0}  \\
\sum_{\ell=0}^\infty \me{\mathrm{final}}{\Tb^{(1)}_\ell}{\pi_0} \\
\vdots \\
\sum_{\ell=0}^\infty \me{\mathrm{final}}{\Tb^{(\nmax)}_\ell}{\pi_0} \\
\end{array}
\right] = 
\left[
\begin{array}{c}
\tbar^{(0)} \\
\tbar^{(1)} \\
\vdots \\
\tbar^{(\nmax)} \\
\end{array}
\right],
\eeq

\noindent where $\tbar^{(n)} = \sum_{\ell=0}^\infty \tbar^{(n)}(\ell) = \Tmom{n}$ is total time moment over all paths.  The matrix $\Fb$ similarly acts on the action vectors $\ket{\eta(\ell)}$ and $\ket{\eta}$:

\beq
\begin{split}
\Fb \ket{\eta(\ell)} & = 
\left[
\begin{array}{c}
\sbar^{(0)}(\ell) \\
\sbar^{(1)}(\ell) \\
\vdots \\
\sbar^{(\nmax)}(\ell) \\
\end{array}
\right] 
 = \ket{\sbar(\ell)}, \\
\Fb \ket{\eta} & = 
\left[
\begin{array}{c}
\sbar^{(0)} \\
\sbar^{(1)} \\
\vdots \\
\sbar^{(\nmax)} \\
\end{array}
\right],
\end{split}
\eeq

\noindent where $\sbar^{(n)}(\ell)$ is the $n$th action moment absorbed in all final states at the $\ell$th jump, and $\sbar^{(n)}$ is the total $n$th action moment.

     Finally, for any function of state $B(\s)$, we can calculate the average value of that function at the $\ell$th intermediate jump.  Define the $N(\nmax+1)$-dimensional row vector

\beq
\bra{B} = 
\left[
\begin{array}{cccc}
\left(\sum_{\s} B(\s) \bra{\s}\right) & \zero & \cdots & \zero \\
\end{array}
\right],
\label{eq:B_def}
\eeq

\noindent where there are $\nmax$ zero row vectors $\zero$, each of length $N$. The row vector $\bra{B}$ acts on $\ket{\tau(\ell)}$ to return the value of $B(\s)$ averaged over the probability distribution across all intermediate states at the $\ell$th jump:

\beq
\innerprod{B}{\tau(\ell)} = \sum_{\s} B(\s) \me{\s}{\Tb^{(0)}_\ell}{\pi_0} = \bar{B}(\ell).
\label{eq:Bl_eval}
\eeq

\noindent For example, if $B(\s) = \th{1}(\s)$, $\bar{B}(\ell)$ tells us the unconditional mean time spent at the $\ell$th intermediate jump.  If $B(\s)$ is set to a position in space corresponding to state $\s$ (for systems that allow embedding of states into physical space), $\bar{B}(\ell)$ is the average position at the $\ell$th intermediate jump, which over all $\ell$ traces the average path of the system.


\subsection{Convergence and asymptotic behavior of path sums}
\label{sec:convergence}

     To numerically calculate the foregoing matrix quantities, we must truncate the sums over path lengths $\ell$ at some suitable cutoff $\Lambda$.  If there are no loops in the network, then the jump matrix $\Qb$ is nilpotent, meaning there is a maximum possible path length $\Lambda$ such that $\Qb^\ell = 0$ for all $\ell > \Lambda$.  In this case all sums converge exactly after $\Lambda$ jumps.  If the network has loops, paths of arbitrarily long length have nonzero probability.  We must then choose a desired precision $\epsilon \ll 1$ and truncate the sums at $\ell = \Lambda$ when
     
\beq
1 - \sum_{\ell=0}^\Lambda \rho(\ell) < \epsilon \quad\mathrm{and}\quad \frac{\tbar^{(\nmax)}(\Lambda)}{\sum_{\ell=0}^\Lambda \tbar^{(\nmax)}(\ell) } < \epsilon.
\label{eq:convergence}
\eeq
     
\noindent The first condition guarantees that the total probability has converged: all remaining paths have total probability less than $\epsilon$.  The second condition indicates that the $\Lambda$th contribution to the maximum moment $\nmax$ is sufficiently small relative to the total moment calculated so far.

    A potential problem with the second convergence condition arises when the state space is periodic, so the final states can only be reached in a number of jumps $\ell$ that is an integer multiple of the periodicity (plus a constant).  For instance, square lattices have a periodicity of 2.  In that case, $\tbar^{(\nmax)}(\ell)$ will alternate between zero and nonzero values as $\ell$ alternates between even and odd values.  To prevent these zero values of $\tbar^{(\nmax)}(\ell)$ from trivially satisfying the second condition in Eq.~\ref{eq:convergence}, we also require that $\tbar^{(\nmax)}(\Lambda)$ be nonzero.  A more subtle problem can arise if there are very low-probability paths with very large contributions to the higher time moments.  For example, one can construct a model where there are extremely long paths with probabilities much smaller than $\epsilon$ but which make arbitrarily large contributions to the total time moments due to the waiting time moments at those states.  The algorithm will satisfy the convergence criteria before these paths are summed and therefore miss their contributions.  This is an extreme example, but in general one may need to reconsider the convergence test depending on the properties of the model at hand.

     How does the cutoff $\Lambda$ depend on the maximum moment $\nmax$?  To address this we must determine the asymptotic behavior of $\tbar^{(n)}(\ell)$ for different $n$. As long as the network has loops, the path length probability distribution $\rho(\ell)$ is asymptotically exponential (Eq.~\ref{eq:asymptotic_rho}; see Appendix~\ref{sec:asymptotic_rho}).  To estimate the asymptotic dependence of the higher time moments on path length, we consider the special case of identical waiting time distributions as in Sec.~\ref{sec:homogeneous}.  Since
     
\beq
\tbar^{(n)}(\ell) = \sum_\varphi \Pcal[\varphi] \Tcal^{(n)}[\varphi] \delta_{\ell,\Lcal[\varphi]},
\eeq

\noindent we can use the approximation $\Tcal^{(n)}[\varphi] \approx (\th{1})^n (\Lcal[\varphi])^n$ from Eq.~\ref{eq:time_moment_approx} (valid when the waiting time distributions are not too dispersed) to obtain

\beq
\begin{split}
\tbar^{(n)}(\ell) & \approx \left(\th{1}\right)^n  \sum_\varphi \Pcal[\varphi] \left(\Lcal[\varphi]\right)^n \delta_{\ell,\Lcal[\varphi]} \\
& = \left(\th{1}\ell\right)^n \rho(\ell) \\
& \sim \left(\th{1}\ell\right)^n e^{-\alpha\ell/\lbar}. 
\end{split}
\label{eq:asymptotic_tbarj}
\eeq

\noindent Since $\ell^n e^{-\alpha\ell/\lbar} = e^{-\alpha\ell/\lbar + n\log\ell}$, the higher moments decay nearly exponentially (up to a logarithmic correction) with the same rate as the probability, set by the mean path length $\lbar$.  We expect this asymptotic behavior to remain valid even when the waiting time distributions are not all the same, as long as the length and time moments are approximately proportional; we will empirically verify this expectation in later examples.

     Although all $\tbar^{(n)}(\ell)$ asymptotically decay with exponential dependence on $\ell$, the logarithmic correction in the exponent shifts the exponential regime toward larger $\ell$ for higher moments; this is why we test convergence on the maximum moment $\nmax$ in Eq.~\ref{eq:convergence}. Indeed, $\tbar^{(n)}(\ell)$ in Eq.~\ref{eq:asymptotic_tbarj} is maximized by $\ell_{\mathrm{max}} = n \lbar/\alpha$, after which exponential decay sets in. Thus we expect scaling for the cutoff to be $\Lambda \sim \nmax \lbar$ to leading order.
     

     The approximate exponential dependence of the moments also enables a convergence scheme more sophisticated than Eq.~\ref{eq:convergence}.  Since we know the asymptotic dependence of all moments will be approximately exponential for long paths, we can simply calculate the moments for path lengths until all $\tbar^{(n)}(\ell)$ have reached their exponential tails, and then fit exponential functions and extrapolate to infer the contributions of the longer paths.  Conceptually, this means that all long path behavior is contained in the statistics of shorter paths, since long paths are simply short paths with many loops~\cite{Sun2006, Manhart2013}. In practice this procedure can help to avoid calculating extremely long paths unnecessarily.


\subsection{Numerical implementation in PathMAN}

     We have implemented the aforementioned matrix formulation in a Python script called PathMAN (Path Matrix Algorithm for Networks), available at \url{https://github.com/michaelmanhart/pathman}, with additional scripts for generating examples and analyzing output.  Figure~\ref{fig:pseudocode} shows the pseudocode.  Since the jump matrix $\Qb$ is typically very sparse, we can store all matrices in sparse formats for efficient storage and computation using SciPy's sparse linear algebra module~\cite{SciPy}.  The script is general enough to treat any CTRW on a finite discrete space given a list of states, their jump probabilities, and at least their first waiting time moments.  The current implementation assumes separable waiting time distributions, but modifying it to run the calculations for non-separable distributions (Appendix~\ref{sec:recursion_derivation}) is straightforward.  The user can specify any path boundary conditions (initial distribution and final states) and functions of state $B(\s)$ to average over.  The script reads all input data from plain text files in a simple format (see GitHub repository for documentation).

\begin{figure}
\begin{center}
\footnotesize{
\begin{tabular}{|p{\columnwidth}|} 
\hline
\textbf{Define} initial distribution $\ket{\pi_0}$ and set of final states $\Sfinal$ \\
\textbf{Define} jump and waiting time matrices $\Qb$, $\Thb^{(n)}$ (Eq.~\ref{eq:Thetaj_def}), $\Qbtilde$ (Eq.~\ref{eq:Qtilde_def}) \\
\textbf{Define} transfer matrices $\Kb$ and $\Gb$ (Eq.~\ref{eq:Kb_Gb_def}) \\
\textbf{Define} final state sum matrix $\Fb$ (Eq.~\ref{eq:Fb_def}) \\
\textbf{Define} row vector $\bra{B}$ (Eq.~\ref{eq:B_def}) for each state function $B(\s)$ \\
\textbf{Define} initial transfer vectors $\ket{\tau(0)} = \ket{\eta(0)} = (\ket{\pi_0}, \zero, \zero, \ldots, \zero)^\mathrm{T}$ (Eq.~\ref{eq:tau_eta_initial}) \\
\textbf{Define} initial cumulative vectors $\ket{\tau} = \ket{\tau(0)}$, $\ket{\eta} = \ket{\eta(0)}$ \\
\\
\textbf{For} $\ell \in \{1, 2, 3, \ldots\}$: \\
\hspace{0.5cm} \textbf{Update} $\ket{\tau(\ell)} = \Kb \ket{\tau(\ell-1)}$ \\
\hspace{0.5cm} \textbf{Update} $\ket{\eta(\ell)} = \Gb \ket{\eta(\ell-1)}$ \\
\hspace{0.5cm} \textbf{Increment} $\ket{\tau}$ by $\ket{\tau(\ell)}$ \\
\hspace{0.5cm} \textbf{Increment} $\ket{\eta}$ by $\ket{\eta(\ell)}$ \\
\hspace{0.5cm} \textbf{Set} $\ket{\tbar(\ell)} = \Fb \ket{\tau(\ell)}$ \\
\hspace{0.5cm} \textbf{Set} $\bar{B}(\ell) = \innerprod{B}{\tau(\ell)}$ \\
\\
\hspace{0.5cm} \textbf{If} $1 - \sum_{\ell'=0}^\ell \rho(\ell') < \epsilon$: \\
\hspace{1cm} \textbf{If} $\tbar^{(\nmax)}(\ell) > 0$ \textbf{and} $\tbar^{(\nmax)}(\ell)/\sum_{\ell'=0}^\ell \tbar^{(\nmax)}(\ell') < \epsilon$: \\
\hspace{1.5cm} \textbf{Break} \\
\\
\textbf{Output} distributions $\ket{\tau}$ and $\ket{\eta}$ of cumulative moments over states \\
\textbf{Output} distributions $\ket{\tbar(\ell)}$ and $\bar{B}(\ell)$ over path lengths \\
\hline
\end{tabular}
}
\end{center}
\caption{Pseudocode for the matrix calculations implemented in PathMAN.}
\label{fig:pseudocode}
\end{figure}

     The rate-limiting step of the algorithm is multiplying the transfer matrices $\Kb$ and $\Gb$ with the vectors $\ket{\tau(\ell-1)}$ and $\ket{\eta(\ell-1)}$ (Fig.~\ref{fig:pseudocode}) to obtain $\ket{\tau(\ell)}$ and $\ket{\eta(\ell)}$, so we use this step to estimate the time complexity of the algorithm.  Assume that each state has an average of $\gamma$ outgoing jumps, so that the jump matrix $\Qb$ has $\gamma N$ nonzero entries.  Each transfer matrix has $(\nmax+1)(\nmax+2)/2$ nonzero blocks (Eq.~\ref{eq:Kb_Gb_def}), yielding approximately $\gamma N(\nmax+1)(\nmax+2)/2$ total nonzero entries in $\Kb$ and $\Gb$. Since we multiply these matrices by the $N(\nmax+1)$-dimensional state vectors at each of the $\Lambda$ total jumps, the algorithm scales as
     
\beq
\mathcal{O}\left( \nmax^2 \gamma N \Lambda \right).
\eeq

\noindent Assuming there are loops in the model (i.e., there is no maximum possible path length), the cutoff $\Lambda$ scales linearly with the mean path length $\lbar$, as well as the maximum moment $\nmax$ as argued in Sec.~\ref{sec:convergence}.  For simple random walks, the mean path length scales as a power of the total number of states:

\beq
\lbar \sim \left\{
\begin{array}{ll}
N^{d_\mathrm{w}/d_\mathrm{f}} & \text{for } d_\mathrm{w} > d_\mathrm{f}, \\
N & \text{for } d_\mathrm{w} \leq d_\mathrm{f}, \\
\end{array}
\right.
\label{eq:lbar_scaling}
\eeq

\noindent where $d_\mathrm{w}$ is the dimension of the random walk and $d_\mathrm{f}$ is the fractal dimension of the space~\cite{Bollt2005, Condamin2007}.  Strictly speaking these scaling relations depend on the boundary conditions (proximity of the initial and final states) and the presence of an energy landscape; the scaling relations in Eq.~\ref{eq:lbar_scaling} are a ``worst-case scenario'' when the landscape is flat and the initial and final states are very far from each other.  Altogether this implies the algorithm will scale as

\beq
\begin{array}{ll}
\mathcal{O}(\nmax^3 \gamma N^{1+d_\mathrm{w}/d_\mathrm{f}}) & \text{for } d_\mathrm{w} > d_\mathrm{f}, \\
\mathcal{O}(\nmax^3 \gamma N^2) & \text{for } d_\mathrm{w} \leq d_\mathrm{f}. \\
\end{array}
\eeq

\noindent Alternative recursive expressions for first-passage time moments on 1D lattices scale as $\mathcal{O}(N^3)$~\cite{Goel1974}, as do general methods for solving the backward equation (a linear system)~\cite{Press1992}; our scaling will be equivalent in the extreme case of a fully-connected network where $\gamma = N-1$.  


\section{Examples}
\label{sec:examples}

     We now illustrate the path ensemble approach on a series of simple examples.


\subsection{1D lattice}

     We first consider a Markov CTRW on a 1D lattice.  Let the lattice have $L$ sites with equal and symmetric transition rates between neighboring sites:
     
\beq
\begin{split}
\me{x + 1}{\Wb}{x} = 1 & \quad \text{for } 1 \leq x < L, \\
\me{x - 1}{\Wb}{x} = 1 & \quad \text{for } 1 < x \leq L, \\
\me{y}{\Wb}{x} = 0 & \quad \text{otherwise}. \\
\end{split}
\label{eq:1D_rates}
\eeq

\begin{figure*}
\centering\includegraphics[scale=1.0]{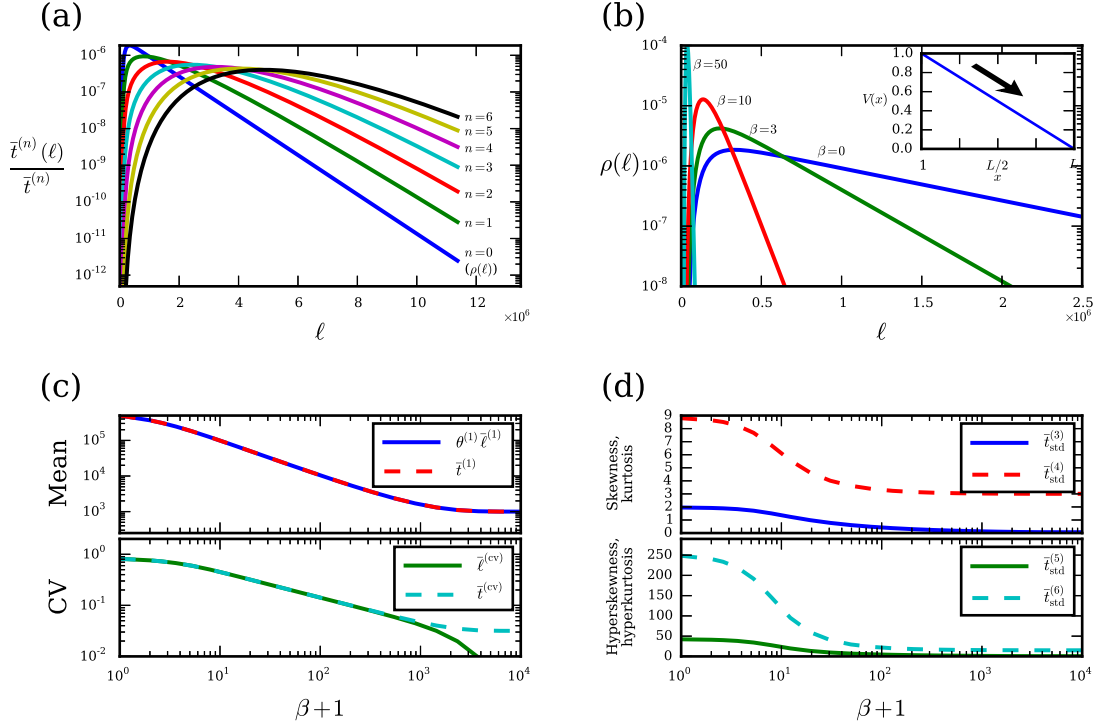}
\caption{
\textbf{Distributions of path lengths and times on a 1D lattice.}  
(a)~The $n$th time moments $\tbar^{(n)}(\ell)$ for paths of length $\ell$, normalized as fractions of the total moments $\tbar^{(n)}$, in the absence of a potential energy.
(b)~Path length probability distribution $\rho(\ell)$ for several choices of $\beta$ on a linear energy landscape $V(x)$ (inset).  
(c)~The mean path length $\lbar^{(1)}$ (scaled by the mean waiting time $\theta^{(1)} = 1/2$ for bulk states), mean path time $\tbar^{(1)}$, path length CV $\lbar^{(\cv)}$, and path time CV $\tbar^{(\cv)}$ as functions of $\beta$.  
(d)~Skewness $\tbar^{(3)}_\std$, kurtosis $\tbar^{(4)}_\std$, hyperskewness $\tbar^{(5)}_\std$, and hyperkurtosis $\tbar^{(6)}_\std$ of path time as functions of $\beta$.  Values of $\beta$ in (c) and (d) are shifted by 1 to show $\beta = 0$ on a log scale.  
In all panels we use a lattice of length $L=1000$ with transition rates given by Eq.~\ref{eq:metropolis_rates}.
}
\label{fig:1Dlattice}
\end{figure*}

\noindent From the rate matrix $\Wb$ we can obtain the jump matrix $\Qb$ and the waiting time moments $\th{n}$ using Eqs.~\ref{eq:Markov_jump_prob} and~\ref{eq:Markov_wtm}; note that the reflecting boundary conditions mean the ``bulk'' states ($1 < x < L$) have $\th{1}_\bulk = 1/2$ while the ``edge'' states ($x=1$, $x=L$) have $\th{1}_\edge = 1$ due to their different connectivities (numbers of outgoing jumps).  We consider the ensemble of first-passage paths from one end of the lattice ($x=1$) to the other ($x=L$).  Figure~\ref{fig:1Dlattice}(a) shows the distributions $\tbar^{(n)}(\ell)$ of path time moments over path lengths; the path length probability distribution $\rho(\ell) = \tbar^{(0)}(\ell)$ is very close to exponential except for small $\ell$, while the higher moments illustrate the Erlang-like function derived in Eq.~\ref{eq:asymptotic_tbarj}.  In particular, we confirm that the higher time moments decay approximately exponentially for large $\ell$.  Since the connectivity is nearly the same everywhere for large $L$ ($\gamma = 2$ for all states except $x=1$ and $x=L$), Eq.~\ref{eq:homogeneous_action} indicates the distribution of path action will also be exponential with mean action (path entropy) $\approx \lbar \log 2$.

     We now introduce a potential energy $V(x) = (L - x)/(L - 1)$ that provides a constant force down the lattice (Fig.~\ref{fig:1Dlattice}(b), inset).  
If we use Metropolis transition rates~\cite{Yeomans1992} $\me{y}{\Wb}{x} = \min(1, e^{-\beta(V(y) - V(x))})$, where $\beta$ is the inverse temperature, we obtain a biased random walk with forward rate of $1$ and backward rate of $e^{-\beta/(L-1)}$:

\beq
\begin{array}{rll}
\me{x + 1}{\Wb}{x} & = 1 & \text{for } 1 \leq x < L, \\
\me{x - 1}{\Wb}{x} & = e^{-\beta/(L-1)} & \text{for } 1 < x \leq L, \\
\me{y}{\Wb}{x} & = 0 & \text{otherwise}. \\
\end{array}
\label{eq:metropolis_rates}
\eeq

\noindent As we increase the inverse temperature $\beta$ from zero, the bias becomes exponentially stronger, leading to a distribution of path lengths more tightly concentrated around the minimum length $\ell = L - 1$ (Fig.~\ref{fig:1Dlattice}(b)).  Since only a single path with probability 1 is available in the limit $\beta \to \infty$, the distributions of path lengths and actions become delta functions at zero; in particular, path entropy is zero because the process is completely deterministic.

     What is the distribution of path times as a function of $\beta$?  In Fig.~\ref{fig:1Dlattice}(c) we show the mean path time $\tbar^{(1)}$, which decreases dramatically as the bias is increased through $\beta$.  It is almost exactly proportional to the mean path length $\lbar^{(1)}$ for all $\beta$, as predicted by Eq.~\ref{eq:time_moment_approx} since $\th{\cv} = 1$ and $\lbar^{(1)} = \lbar \gg 1$.  We also show the coefficients of variation (CVs) $\lbar^{(\cv)}$, $\tbar^{(\cv)}$ of the path length and time distribution, which measure the dispersion.  For $\beta = 0$, both CVs are very close to 1, suggesting that the distributions of path lengths and times are approximately exponential.  However, as $\beta$ becomes large the length CV $\lbar^{(\cv)}$ drops to zero, since the length distribution becomes a delta function, but the time CV $\tbar^{(\cv)}$ decreases to a small but nonzero value, indicating that the distribution becomes narrowly but finitely distributed around its mean. 
     
     Besides CV, standardized moments offer a useful way to characterize the shape of a distribution~\cite{Kendall1994}. They are defined as dimensionless moments of a random variable $X$ shifted and rescaled to have mean 0 and standard deviation 1:

\beq
\average{X^n}_\std = \frac{\average{\left(X - \average{X}\right)^n}}{\left[\average{\left(X - \average{X}\right)^2}\right]^{n/2}}.
\eeq

\noindent Since the first and second standardized moments are 0 and 1 by construction, the lowest non-trivial moment is the third moment, traditionally known as skewness since it measures the asymmetry of the distribution around the mean.  The fourth standardized moment is the kurtosis; the fifth and sixth standardized moments are sometimes called the hyperskewness and hyperkurtosis.
For an exponential distribution, the $n$th standardized moment is $!n$, i.e., the subfactorial or the number of derangements of $n$ objects.  Therefore exponential skewness, kurtosis, hyperskewness, and hyperkurtosis are 2, 9, 44, and 265, respectively.  For a Gaussian distribution, the first four standardized moments are 0, 3, 0, and 15, respectively.

     Figure~\ref{fig:1Dlattice}(d) shows the first four non-trivial ($n \geq 3$) standardized moments of path time on the 1D lattice as functions of $\beta$.  For $\beta = 0$, the standardized moments are very close to their exponential values, confirming that the distribution of first-passage times for a simple random walk is very close to exponential.  
However, as we increase the rightward bias by increasing $\beta$, the moments undergo a rapid transition near $\beta \approx 10$.  Note that this transition happens at rather low temperature ($T = \beta^{-1} = 10^{-1}$) compared to the total change in energy across the lattice, which is set to 1.  For very large $\beta$, the standardized time moments saturate at the Gaussian values of 0, 3, 0, and 15.  This is because a single path of minimal length $\ell = L - 1$ tends to dominate at low temperatures (Fig.~\ref{fig:1Dlattice}(b)), and thus the total path time is just the sum of the waiting times along the single path.  By the central limit theorem, this sum will be approximately Gaussian for large $L$.  Since $\th{1}_\edge = \th{1}_\bulk = \th{2}_{\edge,\c} = \th{2}_{\bulk,\c} = 1$ in this limit (bulk and edge states are the same since travel along the lattice is one-way), the mean and variance of path time in this limit should be the path length $\ell = L - 1 = 999$ times the mean and variance of each waiting time: $\tbar^{(1)} = (L-1)\th{1}_\bulk = 999$ and $\tbar^{(2)}_\c = (L-1)\th{2}_{\bulk,\c} = 999$, leading to a coefficient of variation $\tbar^{(\cv)} = \sqrt{999}/999 \approx 0.03$.  This agrees with Fig.~\ref{fig:1Dlattice}(c).  This simple model is reminiscent of downhill folding in proteins~\cite{Sabelko1999} and linear biochemical pathways such as those used in kinetic proofreading~\cite{Bel2010}, where non-exponential kinetics and the transition between exponential and deterministic (narrow Gaussian distribution) regimes have been previously investigated.  
     
     Equation~\ref{eq:time_moment_approx} suggests that the length and time distributions should be very similar even for $\beta \to \infty$, since the correction term is still small ($\theta^{(\cv)} = 1$ and $\lbar = L - 1 = 999 \gg 1$).  Indeed, the time distribution is a Gaussian narrowly distributed around its mean, whereas the length distribution is a delta function; the errors in the moments are of the order $1/\lbar \approx L^{-1}$.  However, this slight difference is better resolved by considering the complete relation between length and time moments (Eq.~\ref{eq:times_vs_lengths}) with cumulants $\Tmom{n}_\c$ and $\Lmom{n}_\c$ instead of the raw moments.  For $\beta \to \infty$, all $\Lmom{n}_\c = 0$ for $n \geq 2$, which means that we cannot expand Eq.~\ref{eq:times_vs_lengths} for the
cumulants as in Eq.~\ref{eq:time_moment_approx}.  Instead, the only nonzero terms in Eq.~\ref{eq:times_vs_lengths} yield the exact equation $\Tmom{n}_\c = \th{n}_\c \Lmom{}_\c$ for all $n$.
     

%
%


\subsection{1D comb and memory from coarse-graining}

\begin{figure*}
\centering\includegraphics[scale=1.0]{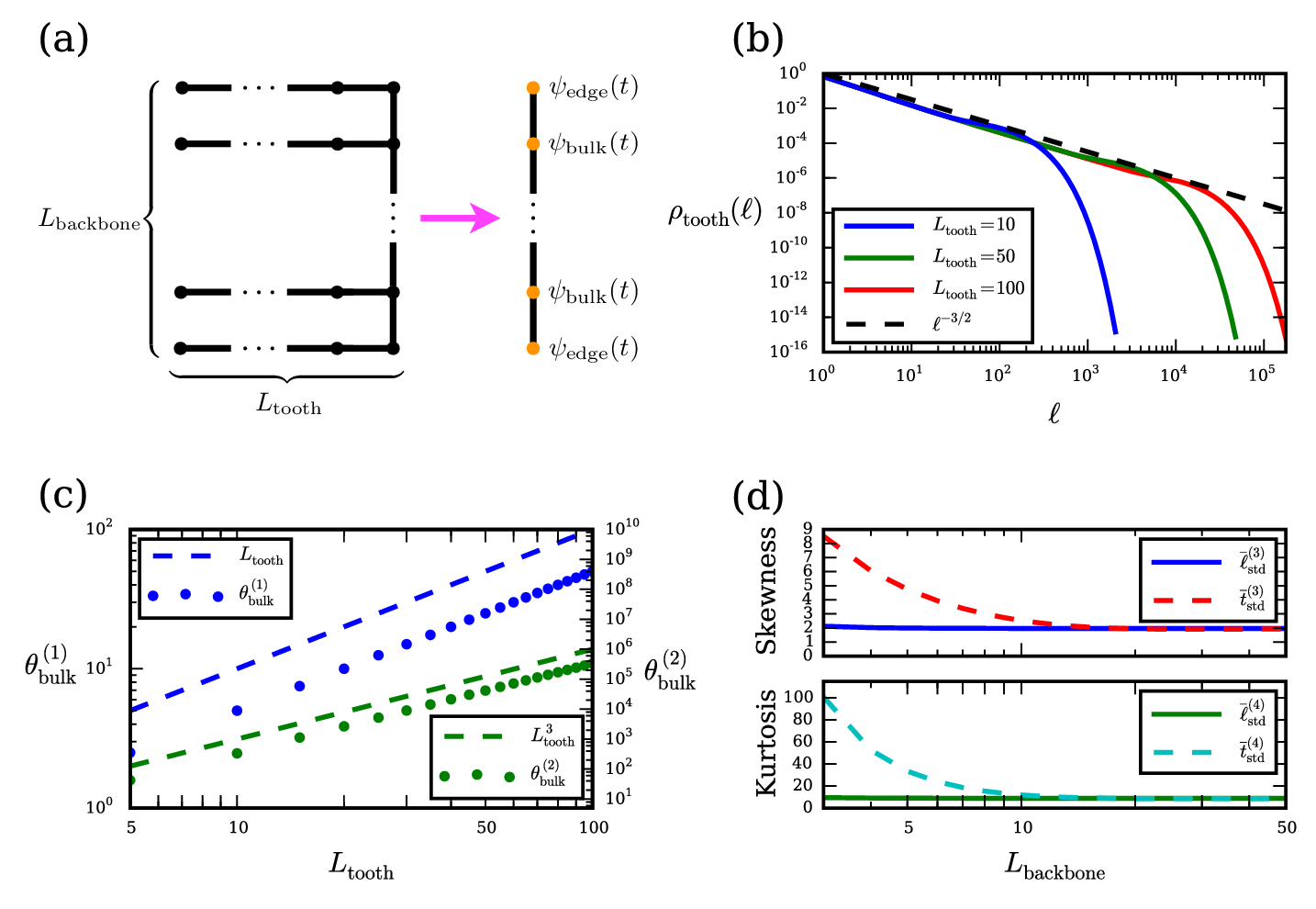}
\caption{
\textbf{Distributions of path lengths and times on a 1D comb.} 
(a)~Schematic of a comb with backbone of length $L_\backbone$ and teeth of length $L_\tooth$.  We coarse-grain the teeth into single states (orange) along the backbone with effective waiting time distributions $\psi_\edge(t)$ and $\psi_\bulk(t)$.  
(b)~Path length distribution $\rho_\tooth(\ell)$ to exit a tooth of different lengths $L_\tooth$, along with the power law $\ell^{-3/2}$ for comparison.  
(c)~Mean $\th{1}_\bulk$ and second moment $\th{2}_\bulk$ of $\psi_\bulk(t)$ as functions of tooth length $L_\tooth$; points are numerical calculations, while dashed lines show expected scaling behavior from Eq.~\ref{eq:moment_scaling}.  
(d)~Skewness $\lbar^{(3)}_\std$, $\tbar^{(3)}_\std$ and kurtosis $\lbar^{(4)}_\std$, $\tbar^{(4)}_\std$ of length and time distributions for paths along the backbone as functions of $L_\backbone$, with $L_\tooth = 100$.
}
\label{fig:comb}
\end{figure*}

     We now turn to an example that explores the effects of waiting memory on distributions of path times.  We consider the 1D comb: a 1D backbone of length $L_\backbone$ where each site has a 1D tooth of length $L_\tooth$ extending from it (Fig.~\ref{fig:comb}(a)).  Combs have traditionally represented simple models of diffusion on percolation clusters and other fractal structures in disordered materials~\cite{Weiss1987, Redner2001}; more recently they have also been proposed as a model for cancer cell proliferation~\cite{Iomin2006}.  As in the previous example (Eq.~\ref{eq:1D_rates}), we use symmetric transition rates of 1 between neighboring sites in the comb.  If we are primarily interested in diffusion along the backbone rather than within the teeth, it is natural to coarse-grain each tooth into a single effective backbone state with some effective waiting time distribution $\psi(t)$ that describes the time spent exploring the tooth before returning to make a jump along the backbone (Fig.~\ref{fig:comb}(a))~\cite{Weiss1987, Redner2001}.  The waiting times within each coarse-grained backbone state are therefore the first-passage times to return to the backbone after exploring the tooth. The distribution of these return times, $f_\tooth(t)$, has the approximate form~\cite{Redner2001}
     
\beq
\psi(t) = f_\tooth(t) \sim \left\{
\begin{array}{ll}
t^{-3/2} & \text{for } t < \tau L_\tooth^2, \\
e^{-t/(\tau L_\tooth^2)} & \text{for } t > \tau L_\tooth^2, \\
\end{array}
\right.
\label{eq:tooth_psi}
\eeq
     
\noindent \noindent where $\tau$ is a time scale that is $\mathcal{O}(1)$ in $L_\tooth$.  Since the distributions of path times and lengths are very similar on 1D lattices in the absence of potential (cf. the $\beta=0$ limit in Fig.~\ref{fig:1Dlattice}(c)), the crossover time $\tau L_\tooth^2$ is essentially the characteristic time scale to explore a 1D lattice of length $L_\tooth$ (Eq.~\ref{eq:lbar_scaling}).  In Fig.~\ref{fig:comb}(b), we show the path length distribution $\rho_\tooth(\ell)$ to exit the tooth, which according to Eq.~\ref{eq:time_moment_approx} should be approximately the same as the distribution of times $f_\tooth(t)$ for large $L_\tooth$.  Indeed, $\rho_\tooth(\ell)$ follows the form of Eq.~\ref{eq:tooth_psi} very clearly: there is a power law regime of $\ell^{-3/2}$ until approximately $\ell \sim L_\tooth^2$, after which there is an exponential decay.  To estimate the tooth waiting time moments $\th{n}$ from $\psi(t) = f_\tooth(t)$, we make the approximation that the power-law regime dominates the moment integrals~\cite{Redner2001}:
     
\beq
\th{n} \sim \int^{\tau L_\tooth^2} dt~ t^{-3/2} t^n \sim L_\tooth^{2n-1}.
\label{eq:moment_scaling}
\eeq

\noindent In Fig.~\ref{fig:comb}(c) we verify this scaling by numerically calculating the moments from first-passage paths that exit the tooth.

     The dominant power-law regime of $\psi(t)$ means that its statistics are very different from those of an exponential distribution.  For example, the CV is $\th{\cv} \sim L_\tooth^{1/2}$ rather than $\sim 1$, indicating a much broader distribution of times compared to the exponential case.  The non-exponential nature of the waiting time distribution is indicative of memory within an effective backbone state: how much longer the system waits in the state depends on how long it has already waited.  Mathematically, this apparent memory arises from coarse-graining each tooth into a single state, which erases information about the position of the system within the tooth. Indeed, for the distribution of times in Eq.~\ref{eq:tooth_psi}, the mean waiting time starting from $t = 0$ is $\sim L_\tooth$, since it is dominated by the power-law regime (Eq.~\ref{eq:moment_scaling}). However, if the system waits at least time $\sim L_\tooth^2$, the mean additional waiting time becomes $\sim L_\tooth^2$, due to the exponential regime. 
In other words, if the system does not leave by the time $\sim L_\tooth^2$ --- meaning that it has diffused far from the backbone --- it is likely to wait much longer as the exponential regime of $\psi(t)$ takes over.

     One effect of this memory is that it can lead to significant differences between the distributions of path times and path lengths along the effective backbone states.  Equation~\ref{eq:time_moment_approx} shows that the moments of path time and path length are approximately proportional if the waiting time distributions are not too broad relative to ratios of path length moments; that is, the correction term in Eq.~\ref{eq:time_moment_approx} is small if $(\th{\cv})^2 \ll \Lmom{n}/\Lmom{n-1}$.  We estimate the size of this correction for the comb model, focusing on first-passage paths from one end of the backbone to the other.  The waiting time CV is $\th{\cv} \sim L_\tooth^{1/2}$ as previously mentioned.  The path length distribution, meanwhile, appears to be very close to exponential: Fig.~\ref{fig:comb}(d) shows that its skewness and kurtosis are consistent with their exponential values (2, 9) for any backbone length.  Since the mean path length should be $\Lmom{} = \lbar \sim L_\backbone^2$ (Eq.~\ref{eq:lbar_scaling}), this implies that the higher moments are $\Lmom{n} \sim n! L_\backbone^{2n}$.  Therefore the correction term in Eq.~\ref{eq:time_moment_approx} is approximately
     
\beq
{n \choose 2} \left(\th{\cv}\right)^2 \frac{\Lmom{n-1}}{\Lmom{n}} \sim \frac{1}{2} (n-1) \frac{L_\tooth}{L_\backbone^2}.
\eeq

\noindent Thus when $L_\tooth \ll L_\backbone^2$, we expect path lengths and times along the backbone to have similar statistics, with a pronounced difference in the opposite limit.  In Fig.~\ref{fig:comb}(d) we calculate skewness and kurtosis of path time moments, varying $L_\backbone$ while fixing $L_\tooth = 100$.  Indeed, for $L_\backbone < \sqrt{L_\tooth} = 10$, there is a large discrepancy between path length and time moments, while for $L_\backbone > 10$ they become very close.  The fact that the length of the teeth must be large compared to the \emph{square} of the backbone length to have an appreciable effect on the path statistics indicates that dynamics along the backbone, rather than within the teeth, tend to dominate the first-passage process. 


\subsection{Memory in coarse-grained metastable states}

     As the previous example showed, memory, in the form of non-exponential distributions of waiting times, naturally arises from coarse-graining because information about the microscopic states of the system is lost. This principle plays a crucial role in the generation of discrete stochastic models for protein folding and other molecular processes~\cite{Prinz2011}.  In these models, a high-dimensional space of ``microscopic'' states (e.g., protein conformations) is coarse-grained into a discrete set of ``macroscopic'' metastable states with some effective transition probabilities.  The resulting coarse-grained model is more amenable to calculating statistical properties of protein dynamics over long time scales, such as the mean folding time or kinetic bottlenecks~\cite{Noe2009, Lane2011}.  However, the coarse-graining can result in qualitative differences between the approximate macroscopic model and the true underlying microscopic dynamics~\cite{Prinz2011},
including non-exponential waiting times in the effective states that are not addressed by conventional transition path theory~\cite{Metzner2009}.
     
\begin{figure*}
\centering\includegraphics[scale=1.0]{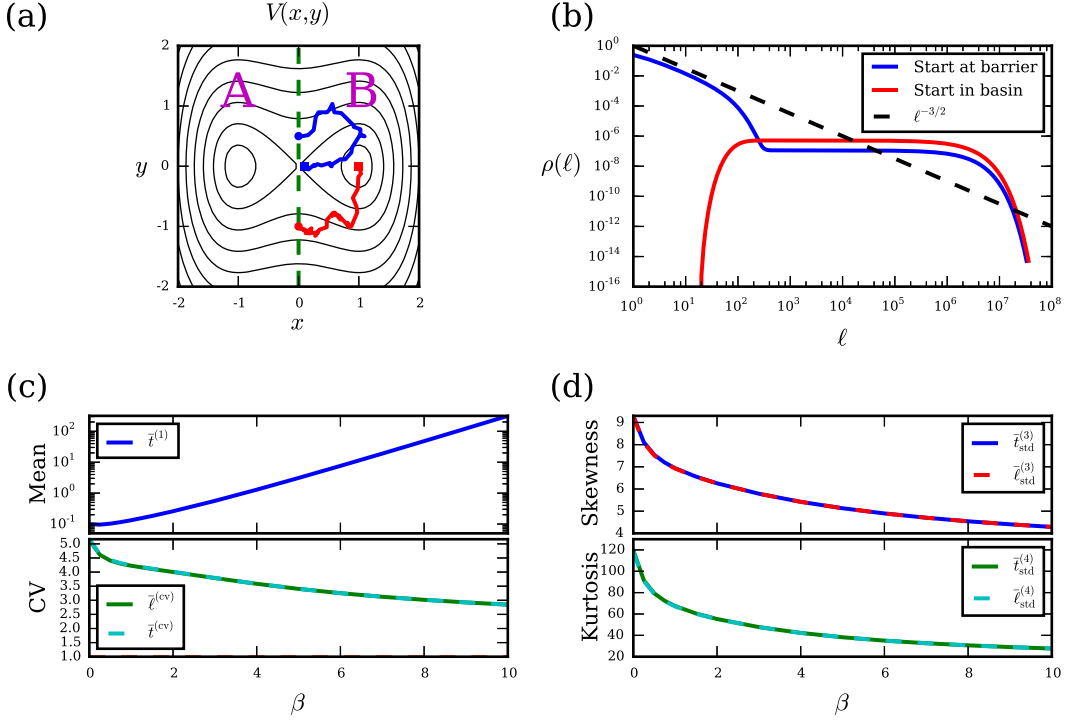}
\caption{
\textbf{Effect of memory in coarse-grained metastable states.} 
(a)~Double-well potential (Eq.~\ref{eq:double_well}) exactly coarse-grained into states A and B with boundary along the green dashed line.  We also show example paths (solid blue and red lines) that both exit B (circles) but start from different initial conditions within B (squares).  
(b)~Length distributions $\rho(\ell)$ of paths (with $\beta = 10$) that exit B but start at different initial conditions (at the central barrier, corresponding to the blue square in (a), or in the low-energy basin, corresponding to the red square in (a)), along with the power law $\ell^{-3/2}$ for comparison.  
For paths that exit the coarse-grained state B, 
(c)~the mean time $\tbar^{(1)}$ and CVs $\lbar^{(\cv)}$, $\tbar^{(\cv)}$ of length and time, and
(d)~skewness $\lbar^{(3)}_\std$, $\tbar^{(3)}_\std$ and kurtosis $\lbar^{(4)}_\std$, $\tbar^{(4)}_\std$ of length and time, all as functions of $\beta$.  
All calculations use a discretized square lattice with $\Delta x = 0.05$ over the space $(x,y) \in [-2, 2] \times [-2,2]$. 
}
\label{fig:barrier_recrossing}
\end{figure*}

     As a simple illustration of this phenomenon, we consider a 2D double-well potential 
     
\beq
V(x, y) = (x^2 - 1)^2 + 2y^2,
\label{eq:double_well}
\eeq

\noindent as shown in Fig.~\ref{fig:barrier_recrossing}(a).  This potential has two local minima at $(\pm 1, 0)$ with $V = 0$, a central barrier at $(0,0)$ with $V = 1$, and reflecting boundaries at $x,y = \pm 2$.
At low temperatures, the system will spend most of its time in the basins around the two minima.  Therefore it is natural to coarse-grain the ``microscopic'' 2D space into two metastable states, A and B, separated by the central energy barrier (Fig.~\ref{fig:barrier_recrossing}(a)).  To characterize the statistics of the two-state dynamics, it is common to calculate a single reaction rate (inverse of the mean time) from one state to another.  However, using a single rate parameter implicitly assumes that waiting times in the coarse-grained states are distributed exponentially. Here we show this to be a poor approximation.

     When the system first transitions to B from A, it starts just to the right of the interface separating the two states (Fig.~\ref{fig:barrier_recrossing}(a)).  The effective waiting time in the coarse-grained state B is therefore the time until the system first returns to that interface, starting from one step off it.  We explicitly calculate the first-passage paths for this microscopic process using our numerical method.  We discretize the space into a 2D lattice with $\Delta x = 0.05$, and we assume a Markov CTRW on the lattice with Metropolis transition rates for jumps between nearest neighbors.  Although the system can enter state B at any point along the interface with A, for simplicity we assume that it enters through the central barrier at $(0,0)$ (the lowest-energy point along the boundary) and therefore starts in B at $(\Delta x, 0)$ (marked by the blue square in Fig.~\ref{fig:barrier_recrossing}(a)).  For low temperature ($\beta = 10$), in Fig.~\ref{fig:barrier_recrossing}(b) we show the distributions of path lengths (blue line) starting from this point and returning anywhere along the interface; an example of such a path is shown in Fig.~\ref{fig:barrier_recrossing}(a) (blue line).  The distribution has a power-law regime for small $\ell$ and exponential regime for large $\ell$.  These two asymptotic limits are the same as the distribution of waiting times in the 1D tooth (Eq.~\ref{eq:tooth_psi}, Fig.~\ref{fig:comb}(b)), and indeed they have a similar physical basis: the power-law regime arises from paths that quickly return to the interface without falling into the low-energy basin, while the exponential regime arises from paths that fall into the low-energy basin before returning.  In contrast to the 1D comb, though, there is a broad flat region of the distribution between the power law and exponential regimes.  This is actually part of the distribution of paths that fell into the low-energy basin before returning; it corresponds to an intermediate regime before that distribution hits its asymptotic exponential regime (cf. path length distributions on a 1D lattice in Fig.~\ref{fig:1Dlattice}(b)).  We confirm this by directly calculating paths with the starting point in the basin (red line in Fig.~\ref{fig:barrier_recrossing}(b); an example path is shown in Fig.~\ref{fig:barrier_recrossing}(a) from the red square to the red circle).

     In Fig.~\ref{fig:barrier_recrossing}(c) and~\ref{fig:barrier_recrossing}(d) we demonstrate that the distribution of path times, i.e., the waiting times in the coarse-grained state B (or A by symmetry), is nearly identical to the path length distribution shown in Fig.~\ref{fig:barrier_recrossing}(b).  Besides the mean time $\tbar^{(1)}$, we also calculate the CV and standardized moments for both length and time distributions, which are indistinguishable over the entire range of $\beta$ (despite the heterogeneities in waiting time distributions across the lattice).  Hence the path length distribution in Fig.~\ref{fig:barrier_recrossing}(b) also describes the effective waiting time distribution in the coarse-grained state.  The power-law regime of this distribution for short paths leaves a distinct signature in the moments.  Even at $\beta = 10$, which represents a temperature that is 10 times smaller than the lowest energy barrier (such that we expect the metastable approximation of A and B to be very good), the distribution of times deviates strongly from an exponential distribution: the CV is nearly 3, while the skewness and kurtosis are much larger than their exponential expectations.  At lower $\beta$ (higher temperatures), the deviation from an exponential distribution becomes even more pronounced.  As with the comb, this enrichment of the distribution for very short paths means that given the system just transitioned to B, it is likely to quickly transition back to A.  But if it does not transition back quickly, it is likely to wait much longer as it falls into the basin and the waiting times become exponentially distributed.
     
     From the comb and double-well examples we can deduce some general principles for the waiting memory that results from coarse-graining the state space.  Assume that the microscopic state space for a system is $d$-dimensional Euclidean space, which we coarse-grain into $N_\macro$ effective macroscopic states, each consisting of $N_\micro$ microscopic states.  The interfaces between coarse-grained states have dimension $d - 1$.  When the system first enters one of these coarse-grained states, it begins just inside an interface.  Therefore the waiting time distribution $\psi(t)$ in the coarse-grained state is the first-passage time to return to that $(d-1)$-dimensional interface.  This return process is effectively a 1D random walk, since only the direction normal to the interface matters (at least within a neighborhood of the initial state, assuming the interface is locally flat).  Therefore the distribution of first-passage times to return to the interface will be the same as for the 1D tooth in the comb (Eq.~\ref{eq:tooth_psi}):
     
\beq
\psi(t) \sim \left\{
\begin{array}{ll}
t^{-3/2} & \text{for } t < \tau N_\micro^\nu, \\
e^{-t/(\tau N_\micro^\nu)} & \text{for } t > \tau N_\micro^\nu, \\
\end{array}
\right.
\label{eq:coarse_grained_psi}
\eeq

\noindent where the crossover time between these regimes is the characteristic time scale $\tau N_\micro^\nu$ to explore the coarse-grained state (Eq.~\ref{eq:lbar_scaling}; $\nu = 2$ for $d = 1$, $\nu = 1$ for $d \geq 2$), and $\tau$ is a microscopic time scale which is ${\cal O} (1)$ in $N_\micro$.  The waiting time moments are approximately

\beq
\th{n} \sim \int^{\tau N_\micro^\nu} dt~ t^{-3/2} t^n \sim N_\micro^{\nu(n-1/2)}.
\eeq

\noindent In particular, the CV is

\beq
\th{\cv} = \frac{\sqrt{\th{2}_\c}}{\th{1}} \sim N_\micro^{\nu/4} = \left\{
\begin{array}{ll}
N_\micro^{1/2} & \text{for } d = 1, \\
N_\micro^{1/4} & \text{for } d \geq 2. \\
\end{array}
\right.
\label{eq:coarse_grained_CV}
\eeq

\noindent As with the 1D comb, the CV scales as a power of the coarse-grained state size, but rather slowly.

     We can now determine whether such waiting time distributions will lead to different statistics of path lengths and times in the coarse-grained model.  Since the mean path length in the coarse-grained model is $\lbar \sim N_\macro^\nu$ (assuming the microscopic and coarse-grained spaces have the same dimensionality), the condition $(\th{\cv})^2 \ll \lbar$ (Eq.~\ref{eq:correction2}) for equivalent path length and time statistics becomes 

\beq
N_\micro \ll N_\macro^2.
\label{eq:coarse_grained_limit}
\eeq
     
\noindent This is consistent with the condition found for the 1D comb where $N_\micro  = L_\tooth$ and $N_\macro = L_\backbone$.  For the double-well model, $N_\micro \approx 3200$ (number of microscopic lattice points in A or B) and $N_\macro = 2$; Eq.~\ref{eq:coarse_grained_limit} does not hold in this case, so we expect significant differences in the statistics of path lengths (jumps between A and B) and path times in the coarse-grained model.  In general, Eq.~\ref{eq:coarse_grained_limit} implies that the more coarse-graining there is (resulting in fewer but larger coarse-grained states), the more significant the memory effects are on the effective CTRW.


\subsection{Random barrier model and memory from spatial disorder}

     To further demonstrate the effects of a complex energy landscape on path statistics, we consider the random barrier model~\cite{Haus1987, benAvraham2000} (RBM), a simple model of transport in disordered systems.     
In this model, a particle diffuses across a regular lattice with quenched energy barriers of random height between neighboring points.  Here we consider a 2D lattice with energy barriers drawn from an exponential distribution 
$p(E) = E_0^{-1}e^{-E/E_0}$, where $E_0$ is the average energy~\cite{Argyrakis1991}.  We assume a Markov CTRW on this lattice with symmetric transition rates between neighboring states that depend exponentially on the intervening energy barrier:
     
\beq
\me{x',y'}{\Wb}{x,y} = \Gamma_0 e^{-\beta E(x',y';x,y)},
\label{eq:RBM_rates}
\eeq
     
\noindent where $\Gamma_0$ is the rate of traversing a barrier of zero height (maximum possible rate), and $E(x',y';x,y)$ is the energy barrier between $(x',y')$ and $(x,y)$.  We use reflecting boundary conditions and set $\Gamma_0 = E_0 = 1$ without loss of generality, as these two quantities set the overall time and energy scales.  
From the rates in Eq.~\ref{eq:RBM_rates} we determine jump probabilities and exponential waiting time moments using Eqs.~\ref{eq:Markov_jump_prob} and~\ref{eq:Markov_wtm}.


\begin{figure*}
\centering\includegraphics[scale=1.0]{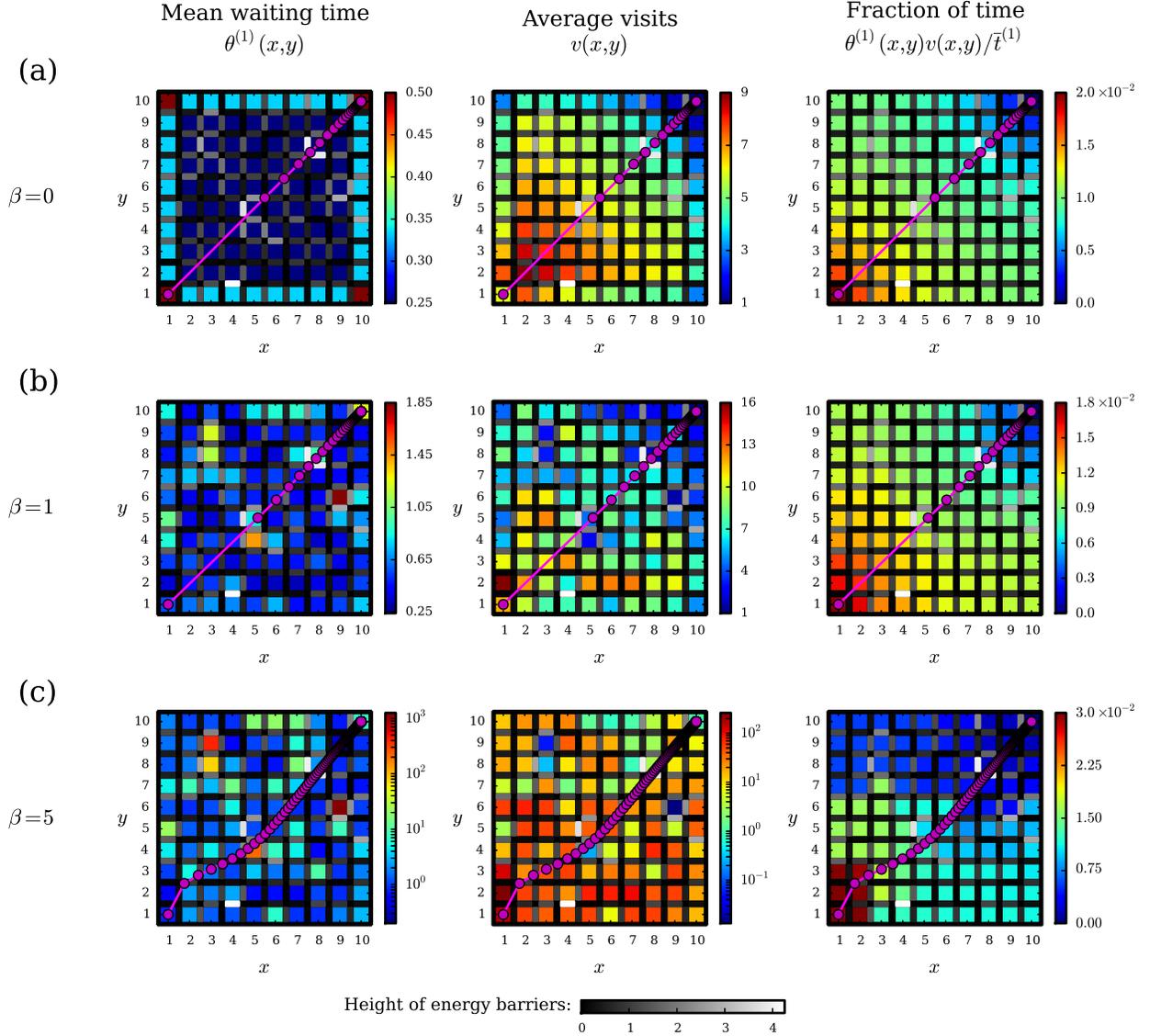}
\caption{
\textbf{Spatial properties of first-passage paths in the random barrier model.}  
For a $10 \times 10$ lattice, we show statistics of first-passage paths from $(1,1)$ to $(10,10)$ for a single quenched realization of the energy barriers.  Each colored cell corresponds to a lattice point $(x,y)$, with gray-scale bars indicating energy barriers between lattice points (higher energies are white, lower energies are black).  Energy barriers are randomly sampled from an exponential distribution with mean $E_0 = 1$, and the transition rate across a zero energy barrier is $\Gamma_0 = 1$.  The leftmost column shows the mean waiting time $\theta^{(1)}(x,y)$, the middle column is the average number of visits $v(x,y)$, and the rightmost column is the average fraction of time $\theta^{(1)}(x,y) v(x,y)/\tbar^{(1)}$ spent at each lattice point. Rows correspond to different inverse temperatures: 
(a)~$\beta = 0$,  
(b)~$\beta = 1$, and 
(c)~$\beta = 5$.  
Magenta points show the average particle position for every 100th jump (connected by straight lines to guide the eye).
}
\label{fig:random_barrier_lattice}
\end{figure*}

     Figure~\ref{fig:random_barrier_lattice} shows a single (quenched) realization of the RBM on a $10 \times 10$ lattice for different values of $\beta$.  
In each panel cells correspond to lattice points while the gray-scale bars between them indicate the height of the intervening energy barriers (same in all panels).  Due to the exponential distribution of energies, most barriers are low (black), with only a few relatively high barriers (white).  We consider the ensemble of first-passage paths on this landscape from $(1,1)$ (bottom-left corner) to $(10,10)$ (top-right corner).  
First we determine path statistics for $\beta = 0$ (Fig.~\ref{fig:random_barrier_lattice}(a)), where all transition rates are equal and the barriers have no effect (Eq.~\ref{eq:RBM_rates}).  
The leftmost panel of Fig.~\ref{fig:random_barrier_lattice}(a) shows the mean waiting time $\th{1}(x,y)$ in each state.  When all transition rates are equal, $\th{1}(x,y)$ depends only on the state's connectivity: the states in the bulk with more neighbors have shorter mean waiting times than do the edge and especially the corner states, which have fewer neighbors.  The middle panel of Fig.~\ref{fig:random_barrier_lattice}(a) shows the average number of visits $v(x,y)$ to each state during the first-passage process (Sec.~\ref{sec:transfer_matrices}).  For $\beta = 0$, the number of visits depends on both the distance to the final state as well as the state's connectivity: edge and corner states with fewer neighbors are visited less often than are bulk states the same distance from the final state.  When we consider the mean fraction of time spent in each state (the product of $\th{1}(x,y)$ and $v(x,y)$, normalized by the total mean path time $\tbar^{(1)}$; rightmost panel of Fig.~\ref{fig:random_barrier_lattice}(a)), the connectivity-dependence largely disappears, so that the fraction of time depends mostly just on the distance to the final state. 


     For $\beta > 0$ the effects of the random energy barriers emerge.  In Fig.~\ref{fig:random_barrier_lattice}(b) and~\ref{fig:random_barrier_lattice}(c) we show path statistics for $\beta = 1$ and $\beta = 5$.  States with large barriers around them acquire significantly longer mean waiting times, leading to a very broad distribution of time scales; at $\beta = 5$, the mean waiting times span three orders of magnitude (Fig.~\ref{fig:random_barrier_lattice}(c), leftmost panel).
However, states with extremely long mean waiting times also tend to have many fewer visits on average (Fig.~\ref{fig:random_barrier_lattice}(b) and~\ref{fig:random_barrier_lattice}(c), middle panels).  This is because the high energy barriers that make these states difficult to exit also make them difficult to enter in the first place.  In contrast with $\beta = 0$, where $v(x,y)$ was determined by both the distance to the final state and the state's local connectivity, for large $\beta$ the average number of visits becomes predominately determined by the state's local properties, i.e., its mean waiting time, rather than its global position on the lattice.  However, the heterogeneity of $\th{1}(x,y)$ and $v(x,y)$ across the lattice nearly vanishes when considering their product, the mean fraction of time (Fig.~\ref{fig:random_barrier_lattice}(b) and~\ref{fig:random_barrier_lattice}(c), rightmost panels): as with $\beta = 0$, the distance to the final state primarily determines the fraction of time spent at a lattice point.  Instead of varying smoothly across states as for $\beta = 0$, though, at $\beta = 5$ the fraction of time appears to have four distinct plateaus on the 2D lattice.  Within each plateau, the particle spends approximately the same fraction of total time at each lattice point.

     Figure~\ref{fig:random_barrier_lattice} also shows the average of all first-passage paths (magenta lines). We calculate the average path by defining state functions for each spatial coordinate as $B_x(x,y) = x$ and $B_y(x,y) = y$, and using Eq.~\ref{eq:Bl_eval} to determine the mean positions $\bar{B}_x(\ell)$ and $\bar{B}_y(\ell)$ as functions of the intermediate jump $\ell$ along a path.  We plot these mean positions for every 100th jump in Fig.~\ref{fig:random_barrier_lattice} to represent the average path of the particle.  For $\beta = 0$, the average path is necessarily symmetric across the diagonal and asymptotically converges toward the final state (Fig.~\ref{fig:random_barrier_lattice}(a)).  For $\beta > 0$, the energy barriers slightly distort the average path at the beginning, but as the path approaches the final state, these asymmetries largely average out (Fig.~\ref{fig:random_barrier_lattice}(b) and~\ref{fig:random_barrier_lattice}(c)).

     We next consider the distributions of path length, time, and action for the RBM.  For $\beta = 0$, all of these distributions are close to exponential in shape, as expected from previous examples.  For example, the length distribution has CV $\lbar^{(\cv)} \approx 0.89$, skewness $\lbar^{(3)}_\std \approx 1.99$, and kurtosis $\lbar^{(4)}_\std \approx 8.95$.  The moments for path time and action are also very close to these values: indeed, Eqs.~\ref{eq:time_moment_approx} and~\ref{eq:homogeneous_action} imply that these distributions should all be very similar since the network is nearly homogeneous.
     
\begin{figure*}
\centering\includegraphics[scale=1.0]{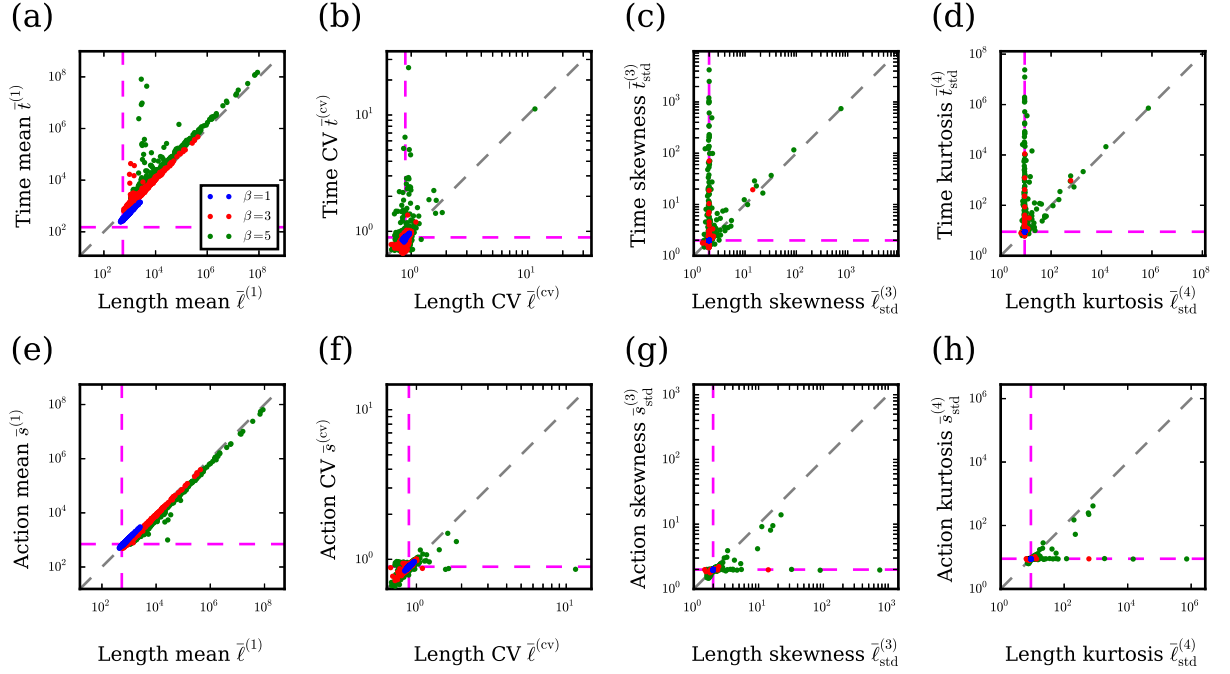}
\caption{
\textbf{Distributions of path statistics in the random barrier model.}  
For 1000 quenched realizations of the energy barriers on a $10 \times 10$ lattice, we show: 
(a)~mean path length $\lbar^{(1)}$ versus mean path time $\tbar^{(1)}$; 
(b)~length CV $\lbar^{(\cv)}$ versus time CV $\tbar^{(\cv)}$; 
(c)~length skewness $\lbar^{(3)}_\std$ versus time skewness $\tbar^{(3)}_\std$; 
(d)~length kurtosis $\lbar^{(4)}_\std$ versus time kurtosis $\tbar^{(4)}_\std$; 
(e)~mean length $\lbar^{(1)}$ versus mean action $\sbar^{(1)}$; 
(f)~length CV $\lbar^{(\cv)}$ versus action CV $\sbar^{(\cv)}$; 
(g)~length skewness $\lbar^{(3)}_\std$ versus action skewness $\sbar^{(3)}_\std$; and 
(h)~length kurtosis $\lbar^{(4)}_\std$ versus action kurtosis $\sbar^{(4)}_\std$.  
Blue points are $\beta = 1$, red points are $\beta = 3$, and green points are $\beta = 5$; the horizontal and vertical dashed magenta lines correspond to the values of the moments for $\beta = 0$.  We also show a diagonal gray line with slope 1 to guide the eye.
}
\label{fig:random_barrier_moments}
\end{figure*}

     What happens to these distributions in the presence of a complex energy landscape  ($\beta > 0$)?  Figure~\ref{fig:random_barrier_moments} shows distributions of the first four moments over many quenched realizations of the RBM at different $\beta$.  For $\beta = 1$ both the mean path length and time are mostly close to their values at $\beta = 0$ (Fig.~\ref{fig:random_barrier_moments}(a)), and their CVs and standardized moments indicate that the distributions are still close to exponential (Fig.~\ref{fig:random_barrier_moments}(b)--(d)).  Larger $\beta$, however, leads to a very wide range of possible length and time moments, which can span several orders of magnitude across realizations. The correlations between path lengths and times are also significant.  Mean lengths and times are mostly clustered along the diagonal, indicating their proportionality for most realizations, but there are some realizations with mean time much larger than mean length (Fig.~\ref{fig:random_barrier_moments}(a)).  For CVs and standardized moments, many realizations that deviate from exponential distributions do so equally in both length and time, resulting in points along diagonal.  However, there are also many realizations with highly non-exponential distributions of path times, even though the length distribution is close to exponential (Fig.~\ref{fig:random_barrier_moments}(b)--(d)).  In this case the approximate equivalence between path length and time in Eq.~\ref{eq:time_moment_approx} breaks down not because $(\th{\cv})^2 \gg \lbar$ --- this is never true in our RBM model since $\th{\cv}(x,y) = 1$ for all $(x,y)$ and $\lbar$ is always large --- but because of the spatial disorder.  Equation~\ref{eq:time_moment_approx} is derived for a network with identical waiting time distributions at all states, but the RBM has a very broad range of mean waiting times for large $\beta$, as the example in Fig.~\ref{fig:random_barrier_lattice}(c) shows.  Thus, a rugged energy landscape even with Markovian waiting times can lead to non-exponential path statistics, and hence the appearance of memory; such non-exponential kinetics have long been discussed in the context of glasses~\cite{Campbell1988, Angelani1998}.

     Figure~\ref{fig:random_barrier_moments}(e)--(h) shows similar distributions of moments for path action (plotted against path length moments for reference).  Equation~\ref{eq:homogeneous_action} shows that the moments of length and action are proportional for networks with homogeneous connectivity.  While the 2D lattice in the RBM is not exactly homogeneous due to boundary conditions, Fig.~\ref{fig:random_barrier_moments}(e) shows mean length and action to be very nearly proportional for almost all realizations.  Since mean action is equivalent to path entropy, its wide range of possible values indicates that first-passage in some realizations is dominated by a few relatively high-probability paths, while in other realizations it is dominated by a large number of much lower-probability paths.  The higher moments of action in Fig.~\ref{fig:random_barrier_moments}(f)--(h) indicate that usually action is nearly exponentially distributed even for larger $\beta$.  In particular, many of the realizations with non-exponential length distributions still have exponentially-distributed actions.  This suggests that the distribution of path actions is much more weakly affected by the energy landscape compared to path lengths and times.


\section{Discussion}
\label{sec:discussion}

     We have studied CTRWs on networks using statistical mechanics of the path ensemble.  A particular convenience of the path formalism lies in exploring the relationship between the distributions of path lengths and path times, which can be viewed as the relationship between the full continuous-time process and its discrete-time projection.  Discrete-time models have generally dominated the theory of random walks not only due to their simplicity, but also because we expect a continuous-time process on the same network to be nearly equivalent under certain conditions~\cite{Weiss1994,benAvraham2000}.  A well-known exception to this expectation is for waiting time distributions $\psi(t)$ without a characteristic time scale (divergent mean), which can produce anomalous diffusion even on regular lattices~\cite{benAvraham2000}. Using our approach, we have identified two more important exceptions.  If all states have identical waiting time distributions $\psi(t)$, Eq.~\ref{eq:time_moment_approx} shows that continuous- and discrete-time dynamics will have different statistics if $(\th{\cv})^2 \gg \Lmom{n}/\Lmom{n-1} \sim \lbar$, where $\th{\cv}$ is the CV of the waiting time distribution $\psi(t)$ and $\lbar$ is the mean path length.  We should therefore expect significant differences between continuous- and discrete-time dynamics to occur when $\psi(t)$ is much more broadly dispersed than an exponential distribution (large $\th{\cv}$) and for small state spaces, which produce small $\lbar$ (Eq.~\ref{eq:lbar_scaling}).  Furthermore, the 2D double-well example suggests this condition is still valid (Fig.~\ref{fig:barrier_recrossing}(d)) even if the waiting time distributions and jump probabilities vary across states, as long as they do not vary too much.  If they do, however, we find another exception to the equivalence of continuous- and discrete-time dynamics: even with exponential waiting times, spatial disorder can produce very different  distributions of path lengths and path times, as illustrated in the random barrier model (Fig.~\ref{fig:random_barrier_moments}).

     Although we have focused primarily on moments of path statistics in this work, ideally we would like to know the entire distributions of these quantities.  In principle one can fit a parameterized distribution to the moments.  In most statistical applications this ``method of moments'' typically produces a good approximation for well-behaved distributions, especially using a very general parameterization such as the Pearson distribution~\cite{Kendall1994}.  For path distributions, a linear combination of exponential functions is likely the most appropriate choice.  Since path length, time, and action distributions are frequently very similar, and since our method explicitly calculates the entire path length distribution already, fitting distributions from moments would be most valuable in cases where the continuous- and discrete-time processes are very different.

     In any case, the moments of path statistics themselves are valuable for quantifying deviations from a simple exponential distribution.  These deviations are important because they represent a form of memory: the amount of time for a process to occur depends on how much time has already passed.  We have emphasized how coarse-graining many ``microscopic'' states of a system into a smaller number of effective ``macroscopic'' states generally leads to non-exponential $\psi(t)$ in the coarse-grained states; we explicitly demonstrated this by coarse-graining teeth in a 1D comb (Fig.~\ref{fig:comb}) and low-energy basins in a double-well potential (Fig.~\ref{fig:barrier_recrossing}).  Furthermore, we have argued that $\psi(t)$ in coarse-grained states will frequently obey Eq.~\ref{eq:coarse_grained_psi}, with a power-law regime for short times and an exponential regime for long times.  Physically, this distribution arises because the system always starts just inside the boundary of a coarse-grained state; therefore, it can either quickly recross the boundary, leading to the power-law regime, or it can explore the rest of the coarse-grained state, leading to the exponential regime.  Compared to a simple exponential distribution, this hybrid distribution is enriched by the power law at short times, meaning that very short waiting times are much more likely than would be expected if the system started in the middle of the coarse-grained state rather than near the boundary.  However, if the system does not quickly exit, it is likely to wait much longer as it explores the rest of the coarse-grained state.  This effective $\psi(t)$ is typically much broader than an exponential distribution, indicated by its larger CV (Eq.~\ref{eq:coarse_grained_CV}); linking this with our condition on path length and time statistics $(\th{\cv})^2 \ll \lbar$, we obtain a condition that shows how much coarse-graining is necessary to see significant memory effects in the statistics of path times (Eq.~\ref{eq:coarse_grained_limit}).
     
     Representing complex state spaces by simpler, coarse-grained representations has long been an implicit element of stochastic models. In recent years it has been explored in Markov models of molecular systems such as proteins~\cite{Noe2009, Lane2011, Prinz2011}.  Non-exponential effects may be important in these systems, especially if the coarse-grained networks are not very large.  Indeed, non-exponential distributions of transition times have previously been found for both protein folding~\cite{Sabelko1999} and enzyme kinetics~\cite{Flomenbom2005b}; Reuveni et al.~\cite{Reuveni2014} showed that these memory effects could lead to qualitatively different properties of enzyme unbinding within the Michaelis-Menten framework.  Our observations underscore the importance of going beyond characterizing such processes by single rates, which implicitly assumes an exponential distribution of times.
     
     Besides waiting memory in the form of non-exponential time distributions, an additional form of memory induced by coarse-graining is in the jump process.  For example, consider a triple-well potential coarse-grained into states A, B, and C.  When the system crosses the barrier from A into B, it is much more likely to jump back into A rather than jump to C, since it begins much closer to A in the microscopic space.  We can account for this in our framework by extending the state space to include not only the current state of the system (e.g., A, B, or C), but also the previous state; the jump process in this extended state space is once again Markovian (although the waiting time distributions remain non-exponential).  We also note that this coarse-graining may require non-separable waiting time distributions $\psi(t|\s\to\s')$, since the distribution of times to return to A from B may be quite different from the distribution to reach C.  Our framework can readily address this generalization (Appendix~\ref{sec:recursion_derivation}).  We look forward to studying the combined roles of jump and waiting memory in coarse-grained molecular models.


\section*{Acknowledgments}

     We thank Pavel Khromov and William Jacobs for careful reading of the manuscript and helpful comments.  M.M. was supported by NIH award F32 GM116217 and A.V.M. was supported by an Alfred P. Sloan Research Fellowship.


\begin{appendix}
\numberwithin{equation}{section}

\section{Asymptotic form of the path length distribution}
\label{sec:asymptotic_rho}

     \red{The path length distribution $\rho(\ell)$ is formally given by Eq.~\ref{eq:rho_def}, which involves the sums of path probabilities $\Pcal[\varphi]$ for all paths $\varphi$ of length $\ell$. More explicitly, we can write $\rho(\ell)$ using matrix elements of powers of the jump matrix $\Qb$ and summing over all final states:}

\beq
\rho(\ell) = \sum_{\s \in \Sfinal} \me{\s}{\Qb^\ell}{\pi_0},
\label{eq:rho_as_matrix_powers}
\eeq

\noindent where $\ket{\pi_0} = \sum_{\s} \pi_0(\s) \ket{\s}$ is the vector of initial state probabilities.  We can decompose $\Qb$ into its Jordan form
     
\beq
\Qb = \mathbf{P}(\mathbf{D} + \mathbf{N})\mathbf{P}^{-1},
\eeq

\noindent where $\mathbf{D}$ is a diagonal matrix with the eigenvalues of $\Qb$, $\mathbf{N}$ is a nilpotent matrix, and $\mathbf{P}$ is an invertible matrix~\cite{Curtis1984}.  Powers of $\Qb$ are therefore

\beq
\begin{split}
\Qb^\ell & = \mathbf{P}(\mathbf{D} + \mathbf{N})^\ell\mathbf{P}^{-1} \\
& = \sum_{\ell'=0}^\ell {\ell \choose \ell'} \mathbf{P} \mathbf{D}^{\ell'} \mathbf{N}^{\ell-\ell'} \mathbf{P}^{-1}. \\
\end{split}
\label{eq:jordan_power}
\eeq

\noindent If $\Qb$ is exactly diagonalizable, then $\mathbf{N} = 0$, and so for large $\ell$, the leading order term in Eq.~\ref{eq:rho_as_matrix_powers} is proportional to $q^\ell = e^{\ell\log q}$, where $q<1$ is the largest eigenvalue of $\Qb$.  If $\Qb$ is not diagonalizable, the leading term will still be proportional to $e^{\ell\log q}$, but may also include a polynomial factor in $\ell$ due to the binomial coefficient in Eq.~\ref{eq:jordan_power}.  However, the polynomial factor only contributes logarithmically to the exponent, i.e., $\ell^k e^{\ell\log q} = e^{\ell\log q + k \log\ell}$, and thus we can neglect it for large $\ell$.  Therefore in general we have $\rho(\ell) \sim e^{\ell\log q}$ for large $\ell$, and since this suggests mean path length must be $\lbar \sim -1/\log q$, we obtain Eq.~\ref{eq:asymptotic_rho}.


\section{Exact relations between path length and time moments using generating functions}
\label{sec:generating_functions}

     \red{Here we derive exact relations between path length and time moments when all states have identical waiting time distributions: $\psi(t|\s) = \psi(t)$.}  We define the moment-generating function for the path length distribution
     
\beq
\tilde{\rho}(s) = \sum_{\ell=0}^\infty \rho(\ell)~ e^{s\ell}, 
\eeq

\noindent so that the moments are

\beq
\Lmom{n} = \tilde{\rho}^{(n)}(0),
\eeq

\noindent where the superscript denotes the derivative:

\beq
\tilde{\rho}^{(n)}(0) = \left.\frac{d^n}{ds^n} \tilde{\rho}(s) \right|_{s=0}.
\eeq

\noindent The cumulant-generating function is therefore $\tilde{\rho}_\c(s) = \log\tilde{\rho}(s)$ with

\beq
\Lmom{n}_\c = \tilde{\rho}_\c^{(n)}(0).
\eeq

\noindent We similarly define the moment- and cumulant-generating functions for the waiting times:

\beq
\begin{array}{lll}
\tilde{\psi}(s) = \displaystyle\int_0^\infty dt~ \psi(t)~ e^{st}, && \th{n} = \tilde{\psi}^{(n)}(0), \\
&& \\
\tilde{\psi}_\c(s) = \log\tilde{\psi}(s), && \th{n}_\c = \tilde{\psi}_\c^{(n)}(0). \\
\end{array}
\eeq

     When the waiting time distributions are $\psi(t)$ for every state, the path time distribution is (Eqs.~\ref{eq:time_dependent_path_prob} and~\ref{eq:path_time_distribution})
     
\begin{multline}
f(t) = \sum_{\ell=0}^\infty \rho(\ell) \int_0^\infty dt_0~ \psi(t_0)~ \int_0^\infty dt_1~ \psi(t_1)~ \cdots \\ 
\times \int_0^\infty dt_{\ell-1}~ \psi(t_{\ell-1})~ \delta\left( t - \sum_{i=0}^{\ell-1} t_i \right).
\label{eq:path_time_distribution_homogeneous}
\end{multline}

\noindent 
Therefore the moment-generating function for path time is

\beq
\begin{split}
\tilde{f}(s) & = \int_0^\infty dt~ f(t)~ e^{st} \\
& = \sum_{\ell=0}^\infty \rho(\ell) (\tilde{\psi}(s))^\ell \\
& = \sum_{\ell=0}^\infty \rho(\ell) e^{\ell \log\tilde{\psi}(s)} \\
& = \tilde{\rho}\left(\tilde{\psi}_\c(s)\right),
\end{split}
\eeq

\noindent while the cumulant-generating function for path time is 

\beq
\begin{split}
\tilde{f}_\c(s) & = \log\tilde{f}(s) \\
& = \log \tilde{\rho}\left(\tilde{\psi}_\c(s)\right) \\
& = \tilde{\rho}_\c\left(\tilde{\psi}_\c(s)\right). \\
\end{split}
\eeq

     We can obtain moments and cumulants of path time by taking derivatives of its generating functions:

\beq
\Tmom{n} = \tilde{f}^{(n)}(0), \quad \Tmom{n}_\c = \tilde{f}_\c^{(n)}(0).
\eeq
     
\noindent To express these in terms of the length and waiting time moments, we use Fa\`{a} di Bruno's formula for derivatives of composite functions~\cite{Comtet1974}:

\begin{multline}
\frac{d^n}{ds^n} g(h(s)) = \sum_{k=1}^n g^{(k)}(h(s)) B_{n,k}(h^{(1)}(s), h^{(2)}(s), \ldots, \\
h^{(n-k+1)}(s)),
\end{multline}

\noindent where $B_{n,k}$ are the partial Bell polynomials and superscripts again denote derivatives. Thus the path time moments are

\beq
\begin{split}
\Tmom{n} =~ & \tilde{f}^{(n)}(0) \\
=~ & \left.\frac{d^n}{ds^n} \tilde{\rho}\left(\tilde{\psi}_\c(s)\right) \right|_{s=0} \\
=~ & \sum_{k=1}^n \tilde{\rho}^{(k)}(\tilde{\psi}_\c(0)) \\
& \times B_{n,k}\left( \tilde{\psi}_\c^{(1)}(0), \tilde{\psi}_\c^{(2)}(0), \ldots, \tilde{\psi}_\c^{(n-k+1)}(0) \right) \\
=~ & \sum_{k=1}^n \Lmom{n} B_{n,k}\left( \th{1}_\c, \th{2}_\c, \ldots, \th{n-k+1}_\c \right). \\
\end{split}
\eeq

\noindent This proves Eq.~\ref{eq:times_vs_lengths}.  We can similarly obtain the path time cumulants:

\beq
\begin{split}
\Tmom{n}_\c =~ & \tilde{f}_\c^{(n)}(0) \\
=~ & \left.\frac{d^n}{ds^n} \tilde{\rho}_\c\left(\tilde{\psi}_\c(s)\right) \right|_{s=0} \\
= & \sum_{k=1}^n \tilde{\rho}_\c^{(k)}(\tilde{\psi}_\c(0)) \\
& \times B_{n,k}\left( \tilde{\psi}_\c^{(1)}(0), \tilde{\psi}_\c^{(2)}(0), \ldots, \tilde{\psi}_\c^{(n-k+1)}(0) \right) \\
= & \sum_{k=1}^n \Lmom{n}_\c B_{n,k}\left( \th{1}_\c, \th{2}_\c, \ldots, \th{n-k+1}_\c \right). \\
\end{split}
\eeq

\section{\red{Approximate relations between path length and time moments based on the central limit theorem}}
\label{sec:central_limit_theorem}

     \red{Here we use the central limit theorem to obtain approximate relations between path length and time moments when all states have identical waiting time distributions: $\psi(t|\s) = \psi(t)$.  Equation~\ref{eq:path_time_distribution_homogeneous} gives the general relation between the length and time distributions in this case, where the nested integrals represent the probability distribution of the sum of the waiting times.}  For sufficiently long paths, this distribution will be approximately Gaussian

\beq
\frac{1}{\sqrt{2\pi\ell\th{2}_\c}} \exp\left( - \frac{\left(t-\ell\th{1}_\c\right)^2}{2\ell\th{2}_\c} \right),
\eeq
 
\noindent and hence

\beq
f(t) \approx \sum_{\ell=0}^\infty \rho(\ell) \frac{1}{\sqrt{2\pi\ell\th{2}_\c}} \exp\left( - \frac{\left(t-\ell\th{1}_\c\right)^2}{2\ell\th{2}_\c} \right).
\eeq

\noindent From this we can obtain approximate relations between the moments.  For example, the first two path time moments are

\beq
\begin{split}
\Tmom{1} & \approx \th{1}_\c \Lmom{}, \\
\Tmom{2} & \approx (\th{1}_\c)^2 \Lmom{2} + \th{2}_\c \Lmom{}, \\
\end{split}
\label{eq:Gaussian_moments}
\eeq

\noindent which are in fact identical to the exact result (Eqs.~\ref{eq:times_vs_lengths} and~\ref{eq:times_vs_lengths_example}) as expected.


\section{\red{Path lengths and times with homogeneous exponential waiting time distributions}}
\label{sec:time_length_moments_exponential}

     When every state has the same exponential waiting time distribution $\psi(t) = \theta^{-1} e^{-t/\theta}$, the sum of the $\ell$ waiting times has an Erlang distribution:
     
\begin{multline}
\int_0^\infty dt_0~ \frac{1}{\theta}e^{-t_0/\theta}~ \int_0^\infty dt_1~ \frac{1}{\theta}e^{-t_1/\theta}~ \cdots \\
\int_0^\infty dt_{\ell-1}~ \frac{1}{\theta}e^{-t_{\ell-1}/\theta}~ \delta\left( t - \sum_{i = 0}^{\ell-1} t_i \right) \\
= \frac{(t/\theta)^{\ell-1}}{(\ell-1)!\theta} e^{-t/\theta}. 
\end{multline}

\noindent Thus the total path time distribution is (using Eq.~\ref{eq:path_time_distribution_homogeneous})

\beq
f(t) = \frac{1}{\theta} e^{-t/\theta} \sum_{\ell=0}^\infty \rho(\ell) \frac{(t/\theta)^{\ell-1}}{(\ell - 1)!}.
\eeq
     
\noindent In this case we can determine the complete distribution of times $f(t)$ given the complete distribution of lengths $\rho(\ell)$.  We can directly calculate the path time moments to be

\beq
\begin{split}
\Tmom{n} & = \int_0^\infty dt~ f(t)~ t^n \\
& = \sum_{\ell=0}^\infty \rho(\ell) \int_0^\infty dt~ \frac{(t/\theta)^{\ell-1}}{\theta(\ell-1)!} e^{-t/\theta} t^n \\
& = \sum_{\ell=0}^\infty \rho(\ell) \theta^n \frac{(n + \ell - 1)!}{(\ell-1)!} \\
& = \theta^n \average{\Lcal(\Lcal + 1) \cdots (\Lcal + n - 1)} \\
& = \theta^n \sum_{k=1}^n \Lmom{k} |s_{n,k}|, \\
\end{split}
\label{eq:times_vs_lengths_exp}
\eeq

\noindent where $|s_{n,k}|$ are the unsigned Stirling numbers of the first kind~\cite{Comtet1974}.  The first few moments are

\beq
\begin{split}
\Tmom{1} & = \theta \Lmom{}, \\
\Tmom{2} & = \theta^2 \left(\Lmom{2} + \Lmom{}\right), \\
\Tmom{3} & = \theta^3 \left(\Lmom{3} + 3\Lmom{2} + 2\Lmom{}\right). \\
\end{split}
\label{eq:times_vs_lengths_exp_example}
\eeq

\noindent This is consistent with the general result in Eq.~\ref{eq:times_vs_lengths} since $\th{j}_\c = (j-1)!~ \theta^j$ for an exponential distribution and $B_{n,k}(0!, 1!, \ldots, (n-k)!) = |s_{n,k}|$~\cite{Comtet1974}.    


\section{Proof of recursion relations for moment matrices}
\label{sec:recursion_derivation}

     We now show that the $\Tb^{(n)}_\ell$ matrices generated by the recursion relation of Eq.~\ref{eq:time_recursion} indeed calculate the path time moments according to Eq.~\ref{eq:recursion_to_time_moments}.  We first successively apply the recursion relation to expand the $\ell$th-order matrix in terms of lower-order matrices:

\begin{widetext}
\beq
\begin{split}
\Tb^{(n)}_{\ell}  & = \Qb \sum_{j_{\ell-1}=0}^n {n \choose j_{\ell-1}} \Thb^{(j_{\ell-1})} \Tb^{(n-j_{\ell-1})}_{\ell-1} \\
& = \Qb \sum_{j_{\ell-1}=0}^n {n \choose j_{\ell-1}} \Thb^{(j_{\ell-1})} \Qb \sum_{j_{\ell-2}=0}^{n-j_{\ell-1}} {n-j_{\ell-1} \choose j_{\ell-2}} \Thb^{(j_{\ell-2})} \Tb^{(n-j_{\ell-1}-j_{\ell-2})}_{\ell-2} \\
& \,\,\, \vdots \\
& = \sum_{j_{\ell-1}=0}^n \sum_{j_{\ell-2}=0}^{n-j_{\ell-1}} \cdots \sum_{j_0=0}^{n - j_{\ell-1} - j_{\ell-2} - \cdots - j_1} {n \choose j_{\ell-1}} {n-j_{\ell-1} \choose j_{\ell-2}} \cdots {n - j_{\ell-1} - j_{\ell-2} \cdots - j_1 \choose j_0} \\
& \quad \times \Qb\Thb^{(j_{\ell-1})} \Qb\Thb^{(j_{\ell-2})} \cdots \Qb\Thb^{(j_0)} \Tb^{(n - j_{\ell-1} - j_{\ell-2} - \cdots - j_0)}_{0} \\
& = \sum_{j_0,j_1,\ldots,j_{\ell-1}} {n \choose j_0, j_1, \ldots, j_{\ell-1}} \Qb\Thb^{(j_{\ell-1})} \Qb\Thb^{(j_{\ell-2})} \cdots \Qb\Thb^{(j_0)},
\end{split}
\label{eq:time_successive_subs}
\eeq
\end{widetext}

\noindent where we have invoked the initial condition $\Tb^{(n - j_{\ell-1} - j_{\ell-2} - \cdots - j_0)}_{0} = \delta_{0,n - j_{\ell-1} - j_{\ell-2} - \cdots - j_0} \mathbf{1}$ from Eq.~\ref{eq:recursion_initial} to obtain the multinomial sum (recall that each summation in the multinomial sum is from $0$ to $n$ subject to the constraint $j_0 + j_1 + \cdots + j_{\ell-1} = n$). Now we take the matrix element of $\Tb^{(n)}_{\ell}$ for the initial distribution $\ket{\pi_0} = \sum_{\s} \pi_0(\s) \ket{\s}$ and $\s_\ell \in \Sfinal$, and insert identities of the form $\sum_\s \ket{\s}\bra{\s}$ to obtain

\begin{widetext}
\beq
\begin{split}
\me{\s_\ell}{\Tb_n^{(\ell)}}{\pi_0} =~ & \sum_{j_0,j_1,\ldots,j_{\ell-1}} {n \choose j_0, j_1, \ldots, j_{\ell-1}} \sum_{\s_0, \s_1, \ldots, \s_{\ell-1}} \me{\s_\ell}{\Qb\Thb^{(j_{\ell-1})}}{\s_{\ell-1}} \me{\s_{\ell-1}}{\Qb\Thb^{(j_{\ell-2})}}{\s_{\ell-2}} \cdots \me{\s_1}{\Qb\Thb^{(j_0)}}{\s_0} \pi_0(\s_0) \\
=~ & \sum_{\s_0, \s_1, \ldots, \s_{\ell-1}} \me{\s_\ell}{\Qb}{\s_{\ell-1}} \me{\s_{\ell-1}}{\Qb}{\s_{\ell-2}} \cdots \me{\s_1}{\Qb}{\s_0} \pi_0(\s_0) \\
& \times \sum_{j_0,j_1,\ldots,j_{\ell-1}} {n \choose j_0, j_1, \ldots, j_{\ell-1}} \th{j_0}(\s_0) \th{j_1}(\s_1) \cdots \th{j_{\ell-1}}(\s_{\ell-1}).  \\
\end{split}
\eeq
\end{widetext}

\noindent Next, we sum over final states $\s_\ell$ and path lengths $\ell$ to obtain

\begin{widetext}
\begin{multline}
\sum_{\ell=0}^\infty \sum_{\s_\ell \in \Sfinal} \me{\s}{\Tb^{(n)}_\ell}{\pi_0} =
\sum_{\ell=0}^\infty \sum_{\s_\ell \in \Sfinal} \sum_{\s_0, \s_1, \ldots, \s_{\ell-1}} \me{\s_\ell}{\Qb}{\s_{\ell-1}} \me{\s_{\ell-1}}{\Qb}{\s_{\ell-2}} \cdots \me{\s_1}{\Qb}{\s_0} \pi_0(\s_0)  \\
\times \sum_{j_0,j_1,\ldots,j_{\ell-1}} {n \choose j_0, j_1, \ldots, j_{\ell-1}} \th{j_0}(\s_0) \th{j_1}(\s_1) \cdots \th{j_{\ell-1}}(\s_{\ell-1}).
\label{eq:time_matrix_element_sum}
\end{multline}
\end{widetext}

\noindent Substituting $\Pcal[\varphi]$ (Eq.~\ref{eq:path_prob}), the time moment functional $\Tcal^{(n)}[\varphi]$ (Eq.~\ref{eq:time_functional}), and the sum over paths

\beq
\sum_\varphi = \sum_{\ell=0}^\infty \sum_{\s_\ell \in \Sfinal} \sum_{\s_0, \s_1, \ldots, \s_{\ell-1}}
\eeq

\noindent into Eq.~\ref{eq:time_matrix_element_sum}, we finally obtain Eq.~\ref{eq:recursion_to_time_moments}:

\beq
\begin{split}
\sum_{\ell=0}^\infty \sum_{\s_\ell \in \Sfinal} \me{\s}{\Tb^{(n)}_\ell}{\pi_0} & = \sum_\varphi \Pcal[\varphi] \Tcal^{(n)}[\varphi] \\
& = \Tmom{n}.
\end{split}
\label{eq:Tmom}
\eeq

     We now show that the recursion relation of Eq.~\ref{eq:general_recursion} are correct for any functional of the form in Eq.~\ref{eq:general_functional}; this will prove the action recursion relation (Eq.~\ref{eq:action_recursion}) as a special case.  As in Eq.~\ref{eq:time_successive_subs}, successive applications of the $\Ub^{(n)}_{\ell}$ recursion relation yield

\begin{multline}
\Ub^{(n)}_{\ell} = \sum_{j_0,j_1,\ldots,j_{\ell-1}} {n \choose j_0, j_1, \ldots, j_{\ell-1}} \\
\times \Ob^{(j_{\ell-1})} \Ob^{(j_{\ell-2})} \cdots \Ob^{(j_0)}.
\end{multline}

\noindent After inserting identities and using the definition $\me{\s'}{\Ob^{(j)}}{\s} = \me{\s'}{\Qb}{\s} (U(\s', \s))^j$, we obtain

\begin{widetext}
\begin{multline}
\me{\s_\ell}{\Ub_n^{(\ell)}}{\pi_0} = \sum_{\s_0, \s_1, \ldots, \s_{\ell-1}} \me{\s_\ell}{\Qb}{\s_{\ell-1}} \me{\s_{\ell-1}}{\Qb}{\s_{\ell-2}} \cdots \me{\s_1}{\Qb}{\s_0} \pi_0(\s_0)\\
\times \sum_{j_0,j_1,\ldots,j_{\ell-1}} {n \choose j_0, j_1, \ldots, j_{\ell-1}} \left(U(\s_\ell,\s_{\ell-1})\right)^{j_{\ell-1}} \left(U(\s_{\ell-1},\s_{\ell-2})\right)^{j_{\ell-2}} \cdots \left(U(\s_1,\s_0)\right)^{j_0}. 
\end{multline}
\end{widetext}

\noindent Just as in Eqs.~\ref{eq:time_matrix_element_sum} and~\ref{eq:Tmom} for time moments, we can then sum over final states and path lengths to show that the $\Ub^{(n)}_{\ell}$ matrices are related to the path ensemble averages via Eq.~\ref{eq:general_functional_moments}.  Finally, we explain how to use the generalized recursion relation in Eq.~\ref{eq:general_recursion} to calculate path time moments when the waiting time distributions are non-separable.  In this case we define $\me{\s'}{\Ob^{(j)}}{\s} = \me{\s'}{\Qb}{\s} \th{j}(\s\to\s')$, where $\th{j}(\s\to\s')$ is the $j$th moment of the non-separable waiting time distribution $\psi(t|\s\to\s')$.  We can then use the recursion relation of Eq.~\ref{eq:general_recursion} to calculate the time moments $\Ub^{(n)}_\ell = \Tb^{(n)}_\ell$.

\end{appendix}

\bibliography{PathMAN_main_text}
\bibliographystyle{unsrt}


\end{document}